\documentclass[journal]{IEEEtran}

\usepackage{color}
\usepackage{cite}
\usepackage{amsmath,amssymb,amsfonts}
\usepackage{array}
\usepackage{makecell}
\usepackage{algorithm,algorithmic}
\usepackage{graphicx}
\usepackage{textcomp}
\usepackage{tikz}
\usetikzlibrary{shapes,arrows}
\usepackage{booktabs}
\usepackage[hidelinks]{hyperref}

\begin{document}

\title{Deep Learning--Assisted UAV Localization Framework for Post-Disaster Search and Rescue Missions}

\author{Xiangjian~Gao,~\IEEEmembership{Member,~IEEE},
        and~Hamid~R.~Sadjadpour,~\IEEEmembership{Senior~Member,~IEEE}%
\thanks{Xiangjian Gao and Hamid R. Sadjadpour are with the Department of
Electrical and Computer Engineering, University of California, Santa Cruz,
Santa Cruz, CA 95064 USA (e-mail: xgao41@ucsc.edu; hamid@ucsc.edu).
\emph{(Corresponding author: Xiangjian Gao.)}}%
\thanks{\copyright~2026 IEEE. Personal use of this material is permitted.
Permission from IEEE must be obtained for all other uses, in any current or
future media, including reprinting/republishing this material for advertising
or promotional purposes, creating new collective works, for resale or
redistribution to servers or lists, or reuse of any copyrighted component of
this work in other works. This is the author's accepted version of the
article published in IEEE Transactions on Aerospace and Electronic Systems,
vol. 62, pp. 10205--10222, 2026, doi: 10.1109/TAES.2026.3687527.}}

\markboth{IEEE Transactions on Aerospace and Electronic Systems}%
{Gao and Sadjadpour: Deep Learning--Assisted UAV Localization for SAR Missions}

\maketitle

\begin{abstract}The precise locating of trapped victims is arguably the most challenging issue in SAR operations, particularly when infrastructure is destroyed and SAR teams only have low-power beacon signals from smartphones to search with. This paper presents a framework for centralized and cooperative UAV-based localization with deep learning–based channel classification and 3D environment-adaptive target estimation. A CNN–LSTM classifier is employed based on a dataset generated at 867.5 MHz with NYUSIM. This classifier labels the individual links of a UAV to a target. Then, these labels activate specific localization solvers: a first-order Taylor-expanded WLS method for LOS settings, an LSRE–SOCP method with iterative refinement for NLOS, and a hybrid projection-based scheme for mixed environments. Through exhaustive simulations, this paper shows that the proposed framework significantly reduces runtime and yields high localization accuracy even when the transmit power and path-loss conditions are unknown. The method can be used in diverse propagation scenarios, making it easy to deploy in reality. The whole NYUSIM-generated dataset is made publicly available to support future research on disaster-aware wireless localization.
\end{abstract}


\begin{table}[t]
\caption{List of Mathematical Notations}
\centering
\renewcommand{\arraystretch}{1.15}
\begin{tabular}{|c|p{5.5cm}|}
\hline
\textbf{Symbol} & \textbf{Definition} \\
\hline
$A$, $\mathbf{b}$ & Linear system matrix and offset vector in WLS \\
$A_\theta$, $A_\varphi$, $A_\beta$ & Azimuth, elevation, and RSS-ratio constraints \\
$\mathbf{a}_k$ & 3D position of the $k$-th UAV \\
$b_k$ & NLOS attenuation bias on link $k$ \\
$b_{\max}$ & Maximum NLOS bias bound \\
$C$, $C_\theta$, $C_\varphi$, $C_{\varphi\theta}$ & Jacobian matrices of first-order perturbations \\
$C_{\beta\theta}$, $C_{\beta\varphi}$, $C_\beta$ & RSS-ratio perturbation Jacobians \\
$C_k$ & Constraint matrix for UAV $k$ in SDP \\
$d_k$ & Euclidean distance between target and UAV $k$ \\
$e_k$ & NLOS bias exponential factor\\
$\eta_k$ & RSS shadowing distortion factor\\
$f$ & Carrier frequency (MHz) \\
$H_{\mathrm{UAV}}$ & UAV altitude \\
$J^{(n)}$ & RSS mismatch cost at iteration $n$ \\
$K$ & Number of UAVs \\
$M$ & Number of RSS samples averaged per link \\
$m_k$, $m_t$ & Scaled received power and transmit power\\
$n^\theta$, $n^\varphi$, $n^s$  & Noise for AOA and RSS measurements\\
$P_t$ & Unknown transmit power of the target \\
$P_{r,k}$ & Received power at UAV $k$ \\
$\text{PL}_{\mathrm{LOS}}$, $\text{PL}_{\mathrm{NLOS}}$ & Path loss in LOS and NLOS portions \\
$Q$ & Covariance matrix of perturbation vector \\
$R_0$ & Coverage radius \\
$s_{kr}$ & SOCP auxiliary variable bounding $\|\zeta \mathbf{a}_k - \chi\|$ \\
$t_{ik}$, $\tau_{ik}$ & SDP/SOCP auxiliary optimization variables \\
$T$ & Total sequence length in LSTM \\
$u_t$ & CNN output feature at LSTM step $t$ \\
$W$ & Covariance of measurement noise \\
$\mathbf{x}$ & 3D target position vector $[x,y,z]^T$ \\
$Z$ & Semidefinite lifting matrix $Z=zz^T$ in SDP \\
$z$ & Lifted vector $z=[\chi^T,\,\zeta]^T$ \\
$\beta_k$ & Path-loss exponent on link $k$ \\
$\chi$, $\zeta$ & Scaled optimization variables ($\chi = \zeta \mathbf{x}$) \\
$\delta_1$ & LOS unit direction vector from UAV 1 \\
$\gamma_1$ & Scalar LOS directional coefficient \\
$\rho$ & Projection coefficient onto LOS ray \\
$\sigma_\theta$, $\sigma_\varphi$, $\sigma_s$ & Standard deviations of AOA/RSS noise \\
$\tau^{(1)}_{kr}$, $\tau^{(2)}_{kr}$ & SOCP rotated-cone variables \\
$\theta_k$, $\varphi_k$ & True azimuth and elevation angles \\
$\varepsilon$ & First-order perturbation vector \\
$\odot$ & Element-wise multiplication\\
\hline
\end{tabular}
\label{tabI}
\end{table}

\section{INTRODUCTION}
\looseness=-1
W{\scshape hen} a major natural disaster—such as an earthquake, wildfire, or flood—strikes, survivors may become trapped beneath collapsed buildings, submerged debris, or within inaccessible terrain, completely cut off from the outside world~\cite{bib1}. Local power grids and cellular base stations are often destroyed or disabled, rendering conventional communication systems inoperable. Meanwhile, rescue personnel and equipment are severely limited, and road networks may be impassable~\cite{bib2}~\cite{bib3}. Recent events, including the July 2025 Central Texas floods and the April 2025 Istanbul earthquake, illustrate how fragile infrastructure can cripple early warning systems and delay critical response efforts. Time is the most precious commodity in such situations: a quicker detection of the survivors with a greater preciseness of their location saves more lives.

Unmanned aerial vehicles (UAVs) can provide a great solution for the post-disaster phase~\cite{bib4}. They can inspect the damaged region from an aerial viewpoint, helping authorities find survivors still trapped under the rubble and act as temporary communication relays when ground-based infrastructure is down~\cite{bib3}. Previous efforts focused on vision-based detection~\cite{bib5}, energy-aware flight planning~\cite{bib6}~\cite{bib7}, reinforcement learning-based swarm coordination~\cite{bib8}. However, such approaches tend to be short of expectations in SAR operations: vision systems in smoke or debris do not function, a trajectory plan relies on earlier maps or predetermined target locations, and a learning-based approach requires broad inputs. More importantly, few existing systems address the core problem of localizing survivors using radio communication systems that are weak, intermittent, and unpredictable.

Given supreme uncertainty in the case of disaster, these conditions may remain for weeks, draining the battery power gradually. With power conservation being their priority, survivors' smartphones occasionally broadcast brief, low-power signals that only contain device IDs. While a little information exists in those broadcasts, they are the only option of tracking it. Since smartphones are usually in as close proximity to the owner as possible, their location is a sufficient indication for providing the survivor's location. Weak signal localization in this power-limited setting, without the usual aids, is fundamentally different from conventional methods.

\looseness=-1
In this regard, we present a comprehensive framework based on a centralized and cooperative approach for SAR operations following a disaster. The fleet of UAVs collects beacons from devices and transmits them to the fusion center for analysis. The fusion center determines and labels line-of-sight (LOS) and non-line-of-sight (NLOS) links using a deep-learning classifier. The overall pattern of the LOS/NLOS is then used to determine the propagation environment. Depending on the result, a localization algorithm is activated either for mixed, NLOS, or purely LOS environments. This technology-embracing approach allows for precise and efficient 3D SAR localizations.

\subsection{Related Literature}
In the aftermath of major natural disasters, victims can end up trapped beneath rubble, submerged in water, or isolated in damaged infrastructure with little or no visibility. Although the Global Positioning System (GPS), widely available on consumer devices, offers meter-level accuracy in LOS environments, its performance degrades sharply in NLOS scenarios like collapsed buildings, smoke, or forests~\cite{bib9}. Also, GPS modules consume considerable energy; activating them can quickly deplete a victim’s limited battery and may trigger background services such as cloud synchronization, further accelerating power loss. Hence, GPS is impractical.

\looseness=-1
Traditional wireless localization is comprised of several different measurement techniques such as time of arrival, time difference of arrival, frequency of arrival, angle of arrival (AoA), and received signal strength (RSS)~\cite{bib10}~\cite{bib11}~\cite{bib12}. However, many of these approaches are unsuitable for post-disaster search and rescue missions. Time-based approaches rely on precise synchronization between transmitters and receivers, which is impossible to realize when the infrastructure is damaged. Frequency-based methods need specific equipment and offer stable frequency performance that is unachievable on most consumer devices. In contrast, AoA and RSS-based localization remain feasible using standard wireless hardware and offer practical solutions. Under LOS conditions, AoA provides trustworthy geometric direction, and RSS is less distorted, enabling accurate and low-complexity estimation. In NLOS environments, however, AoA becomes unreliable due to multipath and reflection, and RSS will fall under severe attenuation and unpredictable bias. In addition, unknown transmit power and path-loss approximation would bring extra challenges, making localization significantly more difficult. Thus, it is vitally important to distinguish between LOS and NLOS conditions before employing any localization algorithm.

\looseness=-1
Machine Learning (ML) and deep learning (DL) algorithms allow for faster and real-time, individually link classification of the environment in LOS/NLOS conditions via RSS channel impulse responses (CIR) properties. Traditional ML methods such as support vector machines (SVM), Gaussian process classifiers (GPC), and random forests (RF) make use of engineering decisions that require calculations, including standard deviation, kurtosis, goodness of fit, log-mean, and the Rician K-factor~\cite{bib13}~\cite{bib14}. Though having the capacity for better functioning in specially designed environments, they have a drawback that they need massive pretraining with the labeled dataset from the environment of deployment. However, in real disaster zones, existing datasets become less applicable because each earthquake or flood area has its own unique signal conditions.

In contrast, DL models automatically learn features from raw input sequences, removing the need for labor-intensive feature engineering and improving their ability to adapt to unseen conditions~\cite{bib15}~\cite{bib16}. Convolutional neural networks (CNNs) are suitable for capturing local spatial or spectral patterns in CIR data. Long short-term memory (LSTM) networks, as a representative recurrent neural network (RNN) model, are well suited for modeling temporal correlations across an entire sequence of signal samples. By combining two methods, a hybrid CNN–LSTM architecture can first extract spatial features via convolution and then process sequential dependencies with recurrent layers to enable robust classification.

In our system, we use a DL classifier to label each UAV/device link as either LOS or NLOS. The fusion center then uses the combined labels from all UAVs connected to a target to infer the overall environment: two or more LOS links indicate a LOS environment, one indicates a mixed environment, and none implies NLOS environment. After identifying the channel condition, the system selects appropriate localization algorithms. For LOS environments, combining RSS and AoA through weighted least squares (WLS) with first-order linearization provides high accuracy at low computational cost. This method is well established and is known to approach the Cramér–Rao lower bound~\cite{bib17}. Although second-order cone programming (SOCP) and semi-definite programming (SDP) can also accommodate unknown transmit power under bounded uncertainty, they typically cause much higher complexity and longer convergence~\cite{bib18}.

In fully NLOS settings, RSS-based localization is faced with severe attenuation and unknown channel parameters. Traditional estimators like weighted least squares (WLS) assume unbiased measurements and perform poorly under obstruction-induced distortion. To improve robustness, prior work has explored robust WLS (RWLS) and least-squares relative error (LSRE) methods~\cite{bib19}~\cite{bib20}~\cite{bib21}, which address additive and multiplicative bias, respectively. However, both approaches suffer from limited accuracy and longer convergence times when jointly estimating the target position, transmit power, and path-loss exponent (PLE) under harsh NLOS conditions. Our prior work~\cite{bib22} proposed an LSRE–SOCP framework that follows a similar iterative refinement strategy but achieves improved accuracy and faster convergence.

In mixed environments, existing approaches like RWLS-SDP, 1AoA–nRSS WLS fusion, and angle reconstruction have been studied~\cite{bib23}~\cite{bib24}~\cite{bib25}. However, angle reconstruction becomes unreliable when distances are derived from NLOS RSS measurements. Thus, we propose a label-aware projection strategy whereby all RSS observations are fed into solver LSRE–SOCP. The links that are NLOS would be subjected to refine iteratively. End result would be the position estimation getting presented on the ray made geometric by the single LOS AOA measurement. This procedure ensures directional consistency. Since LOS PLE is trustworthy, the proposed method accelerates convergence in LSRE–SOCP iterative optimization and avoids instability problems caused by uncertain NLOS angles. The presented architecture produces higher localization accuracy combined with lower runtime than other mixed-environment approaches.

Accurate channel modeling is necessary for evaluating localization performance across large disaster areas. Our prior work introduced a frequency-dependent Friis-based transmission model for through-wall communication by combining ray-tracing with medium-specific losses~\cite{bib26}. While highly effective in indoor or industrial scenarios, this approach depends heavily on detailed channel parameters and structural information, making it infeasible for wide-scale disaster scenarios. Existing UAV air-to-ground channel models also fit poorly here: In LOS cases, the direct path dominates because reflections are mostly absorbed by the ground or surrounding debris; In NLOS cases, detailed modeling is impractical due to the lack of information about obstruction geometry~\cite{bib27}. To overcome these limitations, we utilize information from previous empirical studies that characterize signal attenuation, shadowing, and multipath effects in rubble environments~\cite{bib28}~\cite{bib29}, providing a tractable and expedient modeling approach when detailed geometry is absent.

Existing public datasets for wireless localization are typically limited to narrow use cases: indoor fingerprinting or high-frequency ray-tracing, which do not support outdoor SAR evaluation of RSS/AoA-based and environment-aware localization algorithms~\cite{bib30}~\cite{bib31}. To fill this gap, we generate a dedicated dataset using the NYUSIM simulator in the 866–869 MHz NPSPAC emergency-communication band. The dataset provides co-registered per-link measurements (e.g., RSS, AoA, CIR) together with LOS/NLOS labels, forming a unified resource for both classification and localization tasks. We have released the dataset publicly to support further research in learning-assisted SAR localization.

\subsection{Research Contributions}

Locating individuals after a disaster remains a major challenge for SAR teams. Despite its importance, there are no well-established methods for reliable localization when no communication infrastructure exists and survivor devices emit only lower power and intermittent beacons under highly variable propagation conditions. Prior studies generally assume a single propagation environment, which does not generalize to real SAR scenarios involving heterogeneous link types, unknown transmit power, and uncalibrated PLEs. In addition, most existing frameworks treat channel classification and localization as separate problems and do not incorporate solver selection based on per-link propagation characteristics. This paper addresses these gaps by introducing a unified learning-assisted UAV localization framework that jointly performs per-link classification and environment-adaptive localization across LOS, NLOS, and mixed environments.

\looseness=-1
To meet these challenges, we propose a centralized, cooperative UAV-based localization framework that integrates deep-learning–based per-link channel classification, environment-adaptive localization algorithms, and robust multi-UAV data fusion. This work extends our previous study~\cite{bib22}, which focused on NLOS localization only, into a complete, end-to-end SAR localization pipeline that supports learning-based channel identification, realistic empirical channel modeling, and optimization across various environments. The main contributions are:
\vspace{-2pt}
\begin{itemize}
    \item We design the overall system around post-disaster SAR constraints, including unknown transmit power, uncalibrated PLEs, limited beacon energy, and mixed propagation. To ensure reliable operation, we integrate per-link channel classification, solver selection, and multi-UAV data fusion into a single pipeline.
    
    \item We implement a CNN–LSTM model to perform binary LOS/NLOS classification at the per-link level. The deep-learning method achieves performance comparable to machine-learning classifiers while avoiding manual feature extraction.

    \item After channel classification, the system activates optimized localization solvers for each environment:
    
    \textbf{LOS:} A low-complexity first-order Taylor-based WLS algorithm is used to fuse AOA and RSS measurements to achieve high accuracy.

    \looseness=-1
    \textbf{Mixed:} A label-aware projection strategy is employed in which the LOS link's AOA provides a trusted geometric bearing while RSS measurements from all UAVs are incorporated into an LSRE-SOCP solver. This strategy improves accuracy and accelerates convergence, achieving around $52\%$ lower RMSE and $70\%$ shorter runtime compared with iterative RWLS--SDP in such scenarios (Table~\ref{tabIV}).
    
    \textbf{NLOS:} An enhanced LSRE–SOCP formulation is applied to jointly estimate the target position, transmit power, and PLE under bounded-error assumptions. This robust estimator significantly improves convergence stability and localization accuracy over LSRE–SDP and RWLS–SDP baselines.
    
    \item We construct a comprehensive dataset for UAV-based SAR localization using the NYUSIM v4.0 simulator in the NPSPAC band. The dataset includes both LOS and NLOS propagation to generate diverse and realistic inputs for training and evaluating learning-augmented localization systems. It is publicly released to support broader research and development of post-disaster SAR localization solutions.
\end{itemize}
\vspace{-2pt}

\looseness=-1
This unified framework provides accurate and environment-adaptive localization across diverse environments while remaining compatible with real-world SAR constraints, where accuracy, energy efficiency, and timeliness are essential for saving lives.

The remainder of this paper is organized as follows. Section~II introduces system model, including UAV geometry and empirical propagation under LOS and NLOS conditions. Section~III presents the channel classification framework. Section~IV describes the proposed localization algorithms for various environments. Section~V outlines the simulation setup, dataset generation, and performance evaluation. Section~VI concludes the paper. Table~\ref{tabI} summarizes the notations, and Table~\ref{tabII} lists abbreviations.

\begin{table}[t]
\caption{List of Abbreviations}
\centering
\renewcommand{\arraystretch}{1.15}
\begin{tabular}{|l|l|}
\hline
\textbf{Abbreviation} & \textbf{Definition} \\
\hline
AOA & Angle of Arrival \\
AUC & Area Under the ROC Curve \\
CDF & Cumulative Distribution Function \\
CIR & Channel Impulse Response \\
CNN & Convolutional Neural Network \\
CRLB & Cramér–Rao Lower Bound \\
DL & Deep Learning \\
GPC & Gaussian Process Classifier \\
GPS & Global Positioning System \\
LOS & Line-of-Sight \\
LSRE & Least Squares Relative Error \\
LSTM & Long Short-Term Memory \\
ML & Machine Learning \\
NLOS & Non-Line-of-Sight \\
NPSPAC & National Public Safety Planning Advisory Committee \\
NYUSIM & NYU Wireless Channel Simulator \\
PLE & Path-Loss Exponent \\
ReLU & Rectified Linear Unit \\
RF & Random Forest \\
RMSE & Root Mean Squared Error \\
RNN & Recurrent Neural Network \\
RSS & Received Signal Strength \\
RWLS & Robust Weighted Least Squares \\
SAR & Search and Rescue \\
SDP & Semidefinite Programming \\
SOCP & Second-Order Cone Programming \\
SR-WLS & Significance Regression Weighted Least Squares \\
SVM & Support Vector Machine \\
UAV & Unmanned Aerial Vehicle \\
UMa & Urban Macrocell \\
WLS & Weighted Least Squares \\
\hline
\end{tabular}
\label{tabII}
\end{table}

\section{System Model}
\begin{figure}[h]
\centering
\includegraphics[width=\linewidth]{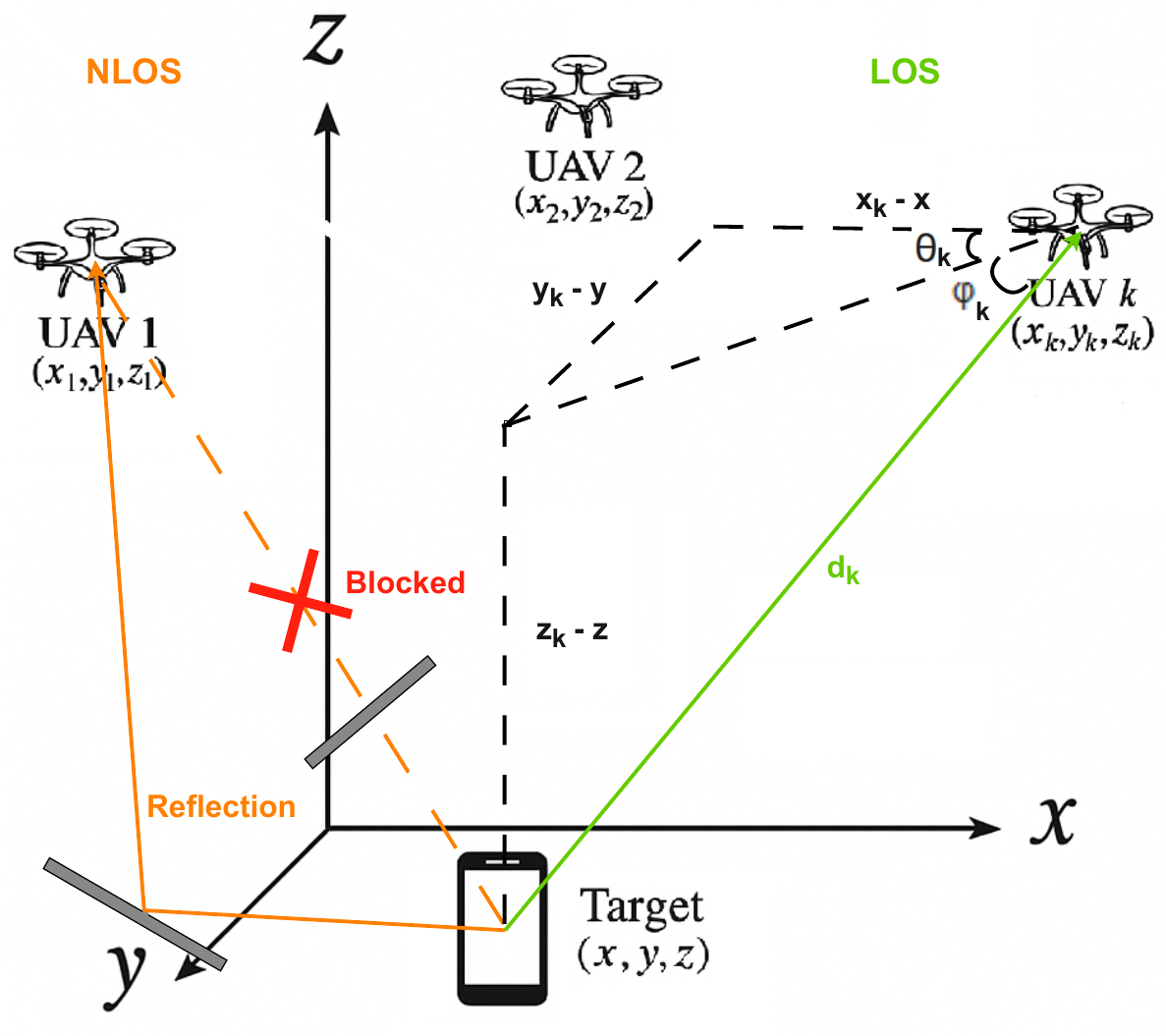}  
\caption{UAV-assisted localization scenario with LOS and NLOS signal propagation.}
\label{fig.1}
\end{figure}

\looseness=-1
In this section, we present the system model for UAV-assisted localization under post-disaster conditions. Fig.~\ref{fig.1} illustrates the overall system geometry and signal propagation environment. We first formulate the measurement models for AOA and RSS, followed by an empirical channel model that capture LOS and NLOS propagation effects in obstructed environments.

\subsection{Angle of Arrival Measurements}

Each UAV is equipped with an antenna array capable of estimating the direction of arrival (AOA) of incoming signals. Let the unknown target location be denoted as $\mathbf{x} = [x, y, z]^T \in \mathbb{R}^3$, and the known position of the $k$-th UAV as $\mathbf{a}_k = [x_k, y_k, z_k]^T$. The true azimuth and elevation angles from UAV $k$ to the target are defined as:
\begin{align}
    \theta_k^0 &= \tan^{-1} \left( \frac{y - y_k}{x - x_k} \right), \label{eq:theta_true} \\
    \varphi_k^0 &= \tan^{-1} \left( \frac{z - z_k}{\sqrt{(x - x_k)^2 + (y - y_k)^2}} \right), \label{eq:phi_true}
\end{align}
where $\theta_k^0 \in (-\pi,\pi]$ and $\varphi_k^0 \in \left(-\frac{\pi}{2},\frac{\pi}{2}\right)$ represent the true geometric azimuth and elevation angles (measured from the horizontal plane), respectively. All angles are in radians unless stated otherwise. The measured angles are corrupted by noise:
\begin{align}
    \theta_k &= \theta_k^0 + n_k^{\theta}, \label{eq:theta_noisy} \\
    \varphi_k &= \varphi_k^0 + n_k^{\varphi}, \label{eq:phi_noisy}
\end{align}
where $n_k^{\theta}$ and $n_k^{\varphi}$ are modeled as independent, zero-mean Gaussian noise with variances $\sigma_\theta^2$ and $\sigma_\varphi^2$, respectively. Let $n^{\theta} = [n_1^{\theta}, \dots, n_K^{\theta}]^T$ and $n^{\varphi} = [n_1^{\varphi}, \dots, n_K^{\varphi}]^T$. Then, the noise vectors satisfy: 
$\mathbb{E}[n^\theta] = 0$, 
$\mathbb{E}[n^\theta (n^\theta)^T] = \sigma_\theta^2 I$, 
$\mathbb{E}[n^\varphi] = 0$, and 
$\mathbb{E}[n^\varphi (n^\varphi)^T] = \sigma_\varphi^2 I$.

\subsection{RSS Modeling under LOS and NLOS Conditions}
Received signal strength is a low-complexity metric for distance estimation in infrastructure-free localization. Let \( P_{r,k} \) denote the received power at the \( k \)-th UAV, where \( k = 1, 2, \ldots, K \). To reduce measurement noise, we average \( M \) temporal samples per link:
\begin{equation}
    P_{r,k} = \frac{1}{M} \sum_{m=1}^{M} P_{r,k}^{(m)},
    \label{eq5}
\end{equation}
where \( P_{r,k}^{(m)} \) is the \( m \)-th received power sample at UAV \( k \), and \( M \) is the total number of samples.

\looseness=-1
Under LOS conditions, the RSS follows a log-normal shadowing model. The received power in linear scale is modeled as:
\begin{equation}
    P_{r,k} = P_t \, d_k^{-\beta_k} \, 10^{\frac{n_k^s}{10}},
    \label{eq6}
\end{equation}
and in logarithmic form as
\begin{equation}
    P_{r,k|\mathrm{dB}} = P_{t|\mathrm{dB}} - 10 \beta_k \log_{10}(d_k) + n_k^s,
    \label{eq7}
\end{equation}
where \(P_t\) is the unknown transmit power of the mobile device, and \(d_k = \| \mathbf{x} - \mathbf{a}_k \|\) is the Euclidean distance between the target and the \(k\)-th UAV. The term \(\beta_k\) denotes the PLE on the \(k\)-th link, and \(n_k^s\) represents zero-mean Gaussian shadowing noise with variance $\sigma_s^2$. The path-loss exponent may vary among UAVs due to differences in geometry, elevation, or local environment.

In NLOS environments, signal propagation is degraded by partial or complete obstruction from rubble, collapsed structures, or debris, resulting in excess attenuation. A widely used RSS-based model is expressed as:
\begin{equation}
    P_{r,k|\mathrm{dB}} = P_{t|\mathrm{dB}} - 10 \beta_k \log_{10}(d_k) - b_k + n_k^s,
    \label{eq8}
\end{equation}
where \(b_k \geq 0\) represents the additional attenuation caused by NLOS obstructions. The bias term \(b_k\) varies with material type, obstacle geometry, and UAV–target orientation, and often dominates localization error in highly obstructed disaster environments.

\subsection{Empirical Channel Modeling in Disaster Scenarios}
To capture realistic signal behavior in obstructed environments, we adopt an empirical path-loss model derived from measurement campaigns in post-disaster scenarios~\cite{bib27}. These studies report extensive propagation measurements at 900 MHz and 1.8 GHz in debris-filled settings, including collapsed buildings and buried transmitters. Among the available frequencies, we select 900 MHz for our simulation due to its strong penetration capability through common building materials and its practical relevance—the adjacent 866–869 MHz band is widely used for national emergency communications.

\looseness=-1
To reflect the physical structure of real NLOS propagation, we partition the total UAV-to-target distance \( d_k \) into two segments: a free-space LOS segment \( d_{\text{LOS}} \), representing the direct path from the UAV to the surface above the buried transmitter, and an obstructed NLOS segment \( d_{\text{NLOS}} \), which penetrates rubble and debris. This decomposition is motivated by the experimental geometry in~\cite{bib24}, where effective signal paths include both open-air and blocked portions. Because UAVs in our system hover tens of meters above ground, well above the test ranges in~\cite{bib27}, we treat this additional altitude as part of the LOS segment, distinct from the empirical NLOS portion.

Accordingly, the total distance is expressed as \( d_k = d_{\text{NLOS}} + d_{\text{LOS}} \), and the total path loss (PL) is approximated by the sum of two components:
\begin{equation}
    \text{PL}_{\text{Total}} = \text{PL}_{\text{LOS}} + \text{PL}_{\text{NLOS}}.
    \label{eq9}
\end{equation}

LOS part is modeled by the Friis free-space equation:
\begin{equation}
    \text{PL}_{\text{LOS}} = 20 \log_{10}(d_{\text{LOS}}) + 20 \log_{10}(f) - 27.55,
    \label{eq10}
\end{equation}
where \( d_{\text{LOS}} \) is the distance in meters for the LOS portion and \( f \) is the carrier frequency in MHz. 

For the obstructed portion, we adopt the log-distance path loss model calibrated from measured data:
\begin{equation}
    \text{PL}_{\text{NLOS}} = L_0 + 10 \beta_k \log_{10}(d_{\text{NLOS}}) + L_{\text{SF}} + L_{\text{SSF}},
    \label{eq11}
\end{equation}
where \( L_0 \) is reference path loss at 1 meter, \( d_{\text{NLOS}} \) is NLOS distance in meters, \( L_{\text{SF}} \) is log-normal shadow fading term, and \( L_{\text{SSF}} \) represents small-scale fading.

 \begin{figure}[h]
\centering
\includegraphics[width=\linewidth]{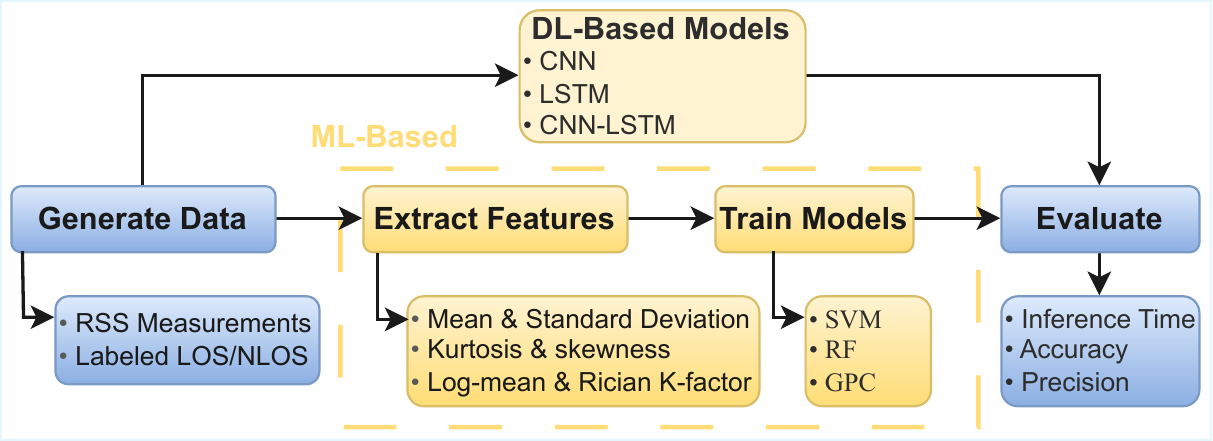}
\caption{Channel classification using ML and DL models (CIR-based).}
\label{fig.2}
\end{figure}

\section{Channel Identification}

The process of determining channel conditions is a crucial step for our proposed system. Distinguishing between LOS and NLOS links allows the system to select appropriate localization algorithms for the observed signal environment. Reliable classification of LOS/NLOS condition is a vital component typical in post-disaster context impacted by unknown physical obstacles and dynamic topology. To solve this, both machine learning (ML) and deep learning (DL) techniques have been proposed using features or sequences extracted from CIR. In this work, we adopt a unified CIR-based framework in which ML classifiers operate on statistical features, while DL models learn directly from raw CIR magnitude sequences. The overall structure of the proposed channel identification framework is illustrated in Fig.~\ref{fig.2}, highlighting the data preprocessing, feature extraction, and classification stages.

\subsection{Machine Learning-Based Classification}

We treat channel identification as a binary task with labels $y\in\{0,1\}$ (LOS $=1$, NLOS $=0$). For algorithms that internally use $\{-1,+1\}$ (e.g., SVM), we define and use $y'^{(i)} \;=\; 2\,y^{(i)} - 1 \;\in\; \{-1,+1\}$, and map predictions back to $\{0,1\}$ via $\hat{y}=(\hat{y}'+1)/2$.

The feature vectors are constructed from CIR magnitude sequences measured over short time windows. From each sequence, statistical features are extracted including the mean and standard deviation (signal stability), skewness and kurtosis (asymmetry and tail behavior), the log-mean (to capture log-normal behavior), the Rician $K$-factor (specular dominance), and goodness-of-fit metrics to theoretical fading models (e.g., Rician or Rayleigh). All features are normalized and then concatenated to form the input vector $\mathbf{p}^{(i)}\in\mathbb{R}^d, i = 1, 2, ..., N$.

\textbf{Support Vector Machine (SVM)} constructs a linear decision boundary in a kernel-induced feature space. After training, the decision function for a test input $\mathbf{p}$ is:
\begin{equation}
\label{eq12}
f(\mathbf{p}) = \sum_{i=1}^{N} \lambda_i \, y'^{(i)} \, k(\mathbf{p}^{(i)}, \mathbf{p}) + w_0,
\end{equation}
where $\lambda_i$ are Lagrange multipliers, $w_0$ is the bias term, and $k(\cdot,\cdot)$ is the kernel. We adopt the RBF kernel in the common $\gamma$ parameterization:
\begin{equation}
\label{eq13}
k(\mathbf{p}, \mathbf{p}') = \exp\!\big( -\gamma \,\|\mathbf{p} - \mathbf{p}'\|^2 \big), \qquad \gamma>0.
\end{equation}
The predicted label in the $\{-1,+1\}$ space is:
\begin{equation}
\label{eq14}
\hat{y}' = \operatorname{sign}\!\big(f(\mathbf{p})\big),
\end{equation}
and is mapped back to $\{0,1\}$ as noted above.
Its inference complexity depends on the number of support vectors and scales as $\mathcal{O}(N_{\mathrm{SV}} \cdot N_f)$, where $N_{\mathrm{SV}}$ is the number of support vectors and $N_f$ is the feature dimension.

\textbf{Gaussian Process Classification (GPC)} is a probabilistic, non-parametric model that places a Gaussian process prior on a latent function $f:\mathbb{R}^d\!\to\!\mathbb{R}$:
\begin{equation}
\label{eq15}
f(\mathbf{p}) \sim \mathcal{GP}\big(0,\, k(\mathbf{p}, \mathbf{p}')\big).
\end{equation}
With a probit link and Laplace approximation, the predictive probability—integrating over the posterior of $f$—is approximated by:
\begin{equation}
\label{eq16}
P(y=1 \mid \mathbf{p}) \;\approx\; \Phi\!\left( \frac{\mu_f(\mathbf{p})}{\sqrt{1 + v_f(\mathbf{p})}} \right),
\end{equation}
where $\mu_f(\mathbf{p})$ and $v_f(\mathbf{p})$ are the posterior mean and variance of $f(\mathbf{p})$, and $\Phi(\cdot)$ is the standard normal cumulative distribution function (CDF). The decision is:
\begin{equation}
\label{eq17}
\hat{y} \;=\; 
\begin{cases}
1, & P(y=1 \mid \mathbf{p}) \ge 0.5,\\
0, & \text{otherwise}.
\end{cases}
\end{equation}
We use a composite kernel combining nonlinear and linear components:
\begin{equation}
\label{eq18}
k(\mathbf{p}, \mathbf{p}') \;=\; \kappa_{\mathrm{rbf}}^{2} \exp\!\left( -\frac{\|\mathbf{p} - \mathbf{p}'\|^2}{2\ell^{2}} \right) \;+\; \kappa_{\mathrm{lin}}^{2} \,\mathbf{p}^\top \mathbf{p}',
\end{equation}
where $\kappa_{\mathrm{rbf}}^{2}$ is the RBF amplitude, $\ell$ is the length scale, and $\kappa_{\mathrm{lin}}^{2}$ scales the linear term to promote correlation.
GPC relies on covariance matrix inversion, resulting in $\mathcal{O}(N_s^3)$ training complexity and $\mathcal{O}(N_s^2)$ inference complexity, where $N_s$ denotes the number of training samples.

\textbf{Random Forests (RF)} improve generalization and robustness to noisy features by combining an ensemble of uncorrelated decision trees at low computational cost. Each tree $h_t:\mathbb{R}^d\!\to\!\{0,1\}$ is trained on bootstrapped samples with random feature subsets and independently produces a prediction for input $\mathbf{p}$. The final decision is by majority vote:
\begin{equation}
\label{eq19}
\hat{y} = \operatorname{mode}\!\left\{ h_t(\mathbf{p}) \right\}_{t=1}^T,
\end{equation}
where $T$ is the number of trees and $\operatorname{mode}(\cdot)$ returns the most frequent output label among all trees. \\
Tree splits minimize the Gini impurity:
\begin{equation}
\label{eq20}
\mathrm{Gini} = 1 - \sum_{c=1}^{\mathcal{N}} p_c^2,
\end{equation}
where $p_c$ is the class proportion at a node and $\mathcal{N}$ is the number of classes.
Its inference complexity scales as $\mathcal{O}(N_t \cdot H_{tree})$, where $N_t$ denotes the number of decision trees and $H_{tree}$ is the average depth of each tree.

\looseness=-1
While SVM, GPC, and RF can separate LOS and NLOS patterns using engineered features, their performance depends on the stability of those feature distributions between training and deployment. In dynamic post-disaster environments, this assumption is fragile, motivating DL models that learn directly from raw CIR sequences without environment-specific feature design.

\subsection{Deep Learning-Based Classification}

The CNN–LSTM architecture integrates spatial and temporal modeling for robust channel classification under nonstationary propagation. Each deployment may vary in terrain, building materials, user geometry, and dynamics, leading to distinct CIR characteristics. DL offers a flexible alternative by learning discriminative representations from raw CIR magnitudes, eliminating manual feature design.

\textbf{Convolutional Neural Networks (CNNs)} extract local patterns from input sequences using shared convolutional filters. Let the $p$-th packet input be $\mathbf{s}_p \in \mathbb{R}^{N}$ CIR magnitude vector. A 1D convolution with kernel width $Q$ and stride $S$ computes:
\begin{equation}
\label{eq21}
z^{[L]}_i \;=\; a\!\left( \sum_{q=0}^{Q-1} w_q^{[L]} \, \mathbf{s}_{i + q \cdot S} \;+\; b^{[L]} \right),
\end{equation}
where $i$ is the output index, $q$ indexes the filter window, $a(\cdot)$ is the activation (e.g., Rectified Linear Unit (ReLU)), and $w^{[L]}_{q}, b^{[L]}$ are the trainable filter and bias at layer $L$. CNN filters capture localized patterns such as fading dips or spectral notches indicative of NLOS.

\textbf{Long Short-Term Memory (LSTM)} networks model temporal dependencies across packets. 
At each time step $t$, the LSTM processes an input feature $\mathbf{u}_t$ (here, $\mathbf{u}_t=\mathbf{z}_t$ from the CNN) and updates its internal state:
\begin{align}
\mathbf{f}_t &= \sigma(W_f \mathbf{u}_t + U_f \mathbf{h}_{t-1} + \mathbf{b}_f), \\
\mathbf{i}_t &= \sigma(W_i \mathbf{u}_t + U_i \mathbf{h}_{t-1} + \mathbf{b}_i), \\
\tilde{\mathbf{c}}_t &= \tanh(W_c \mathbf{u}_t + U_c \mathbf{h}_{t-1} + \mathbf{b}_c), \\
\mathbf{c}_t &= \mathbf{f}_t \odot \mathbf{c}_{t-1} + \mathbf{i}_t \odot \tilde{\mathbf{c}}_t, \\
\mathbf{o}_t &= \sigma(W_o \mathbf{u}_t + U_o \mathbf{h}_{t-1} + \mathbf{b}_o), \\
\mathbf{h}_t &= \mathbf{o}_t \odot \tanh(\mathbf{c}_t),
\end{align}
where $\sigma(\cdot)$ is the logistic sigmoid, $\tanh(\cdot)$ the hyperbolic tangent, and $\odot$ denotes elementwise multiplication. 
The vectors $\mathbf{f}_t$, $\mathbf{i}_t$, and $\mathbf{o}_t$ represent the forget, input, and output gates, respectively. 
The candidate memory state is $\tilde{\mathbf{c}}_t$, the cell state is $\mathbf{c}_t$, and the hidden state is $\mathbf{h}_t$. 
Each gate uses its own learnable matrix parameters $\{W_\ast,U_\ast\}$ and bias vector $\mathbf{b}_\ast$; unless otherwise stated, $\mathbf{h}_0=\mathbf{0}$ and $\mathbf{c}_0=\mathbf{0}$.

\textbf{CNN–LSTM hybrid architecture} combines the spatial feature extraction capability of CNNs with the temporal modeling strength of LSTMs. 
For each CIR packet $\mathbf{s}_p$, the CNN produces a latent embedding $\mathbf{z}_p$, forming the sequence $\{\mathbf{z}_1,\dots,\mathbf{z}_P\}$.
This sequence is fed into the LSTM in order (one packet per time step, $\mathbf{u}_t=\mathbf{z}_t$), which captures inter-packet dynamics. 
After processing all packets (i.e., $T=P$), the final hidden state $\mathbf{h}_T$ aggregates the temporal information and is projected to a binary probability via
\begin{equation}
\label{eq28}
\hat{y} = \sigma(\mathbf{w}^\top \mathbf{h}_T + b),
\end{equation}
where $\sigma(\cdot)$ denotes the sigmoid activation function, and the output-layer parameters $\mathbf{w}$ (vector) and $b$ (scalar) are learnable. 
The model is trained using the binary cross-entropy loss
\begin{equation}
\label{eq29}
\mathcal{L} = -\frac{1}{N} \sum_{i=1}^{N}
\left[ y^{(i)} \log_{10} \hat{y}^{(i)} + \big(1 - y^{(i)}\big) \log_{10}\!\big(1 - \hat{y}^{(i)}\big) \right],
\end{equation}
where $N$ is the number of training samples and $y^{(i)} \in \{0,1\}$ is the ground-truth label.

\section{Proposed Localization Algorithm}
Based on the per‑link LOS/NLOS labels produced by the channel classifier (Section~III), the system activates one of three environment‑adaptive localization algorithms. Each solver is designed to match the specific measurement behavior of its propagation condition, ensuring that the estimator leverages the most reliable information available. The following subsections detail the LOS, NLOS, and mixed‑environment methods, along with their underlying derivations and noise modeling. A high-level overview of the solver-selection process is provided in Algorithm~\ref{Algorithm1}.

\subsection{LOS Environment Localization Algorithm}
\looseness=-1
This RSS- and AOA-based localization algorithm estimates the target's 3D position under LOS conditions using a cooperative weighted least squares (WLS) approach. To handle the unknown transmit power \( P_t \), we adopt a reformulation inspired by~\cite{bib17}, which eliminates \( P_t \) by taking the ratio of RSS measurements between UAV \( k \) and a reference UAV (assumed to be UAV~1). Under LOS conditions, we assume the path-loss exponents \( \beta_k \) are approximately equal across all UAV–target links, i.e., \( \beta_k \approx \beta_1 \). Taking the ratio of received powers between UAV \( k \) and the reference, the distance ratio is:
\begin{equation}
\frac{d_k}{d_1} 
= \left( \frac{P_{r,1}}{P_{r,k}} \right)^{\frac{1}{\beta_k}} 
  \exp\left( \frac{(n_k^s - n_1^s)\ln(10)}{10 \beta_k} \right),
\label{eq30}
\end{equation}
which can be approximated using a first-order Taylor expansion when \( |n_k^s - n_1^s| \) is small:
\begin{equation}
\frac{d_k}{d_1} 
\approx \left( \frac{P_{r,1}}{P_{r,k}} \right)^{\frac{1}{\beta_k}} 
\left[ 1 + \frac{(n_k^s - n_1^s)\ln(10)}{10 \beta_k} \right].
\label{eq31}
\end{equation}

Let:
\begin{equation}
\beta_{1k} \;=\; \left( \frac{P_{r,1}}{P_{r,k}} \right)^{\frac{1}{\beta_k}},
\qquad
\xi \;=\; \frac{\ln 10}{10\,\beta_k},
\label{eq32}
\end{equation}
so that:
\begin{equation}
\frac{d_k}{d_1}
= \beta_{1k}
  + 
  \beta_{1k} \, \xi \, (n_k^s - n_1^s).
\label{eq33}
\end{equation}

In the absence of measurement noise, the true distances satisfy:
\begin{equation}
d_k^0 = d_1^0 \, \beta_{1k}^0,
\quad k = 2, 3, \ldots, K,
\label{eq34}
\end{equation}
where \( \beta_{1k}^0 \) denotes the ideal power ratio component in noise-free conditions.

To relate the azimuth and elevation to the target’s position and UAV coordinates, the angle equations are pseudo-linearized as:
\begin{equation}
\sin\theta_k^0 \,(x - x_k)
- \cos\theta_k^0 \,(y - y_k) = 0,
\label{eq35}
\end{equation}
\begin{equation}
\begin{gathered}
(x - x_k) \sin{\varphi_k^0} \cos{\theta_k^0} + (y - y_k) \sin{\varphi_k^0} \sin{\theta_k^0} \\
- (z - z_k) \cos{\varphi_k^0} = 0.
\end{gathered}
\label{eq36}
\end{equation}

From the geometry in Fig.~\ref{fig.1}, the LOS direction relationships yield:
\begin{equation}
\begin{gathered}
d_k^0 \cos{\varphi_k^0} \cos{\theta_k^0} = x - x_k, \\
d_k^0 \cos{\varphi_k^0} \sin{\theta_k^0} = y - y_k, \\
d_k^0 \sin{\varphi_k^0} = z - z_k.
\end{gathered}
\label{eq37}
\end{equation}

and equivalently:
\begin{equation}
\begin{aligned}
d_k^0=
&(x - x_k)\,\cos\varphi_k^0\,\cos\theta_k^0
 + \\ 
&(y - y_k)\,\cos\varphi_k^0\,\sin\theta_k^0 
 + (z - z_k)\,\sin\varphi_k^0.
\end{aligned}
\label{eq38}
\end{equation}

Substituting the noise-free distance ratio \eqref{eq34} into \eqref{eq38} gives the constraint:
\begin{equation}
\begin{gathered}
\beta_{1k}^0 (x - x_1) \cos{\varphi_1^0} \cos{\theta_1^0} + \beta_{1k}^0 (y - y_1) \cos{\varphi_1^0} \sin{\theta_1^0} \\
+ \beta_{1k}^0 (z - z_1) \sin{\varphi_1^0} 
= (x - x_k) \cos{\varphi_k^0} \cos{\theta_k^0} \\
+ (y - y_k) \cos{\varphi_k^0} \sin{\theta_k^0} 
+ (z - z_k) \sin{\varphi_k^0}.
\end{gathered}
\label{eq39}
\end{equation}

Stacking the angle and RSS-ratio constraints gives a linear system:
\begin{equation}
A \mathbf{x} - \mathbf{b} = 0,
\label{eq40}
\end{equation}
with
\begin{equation}
\begin{gathered}
A = \begin{bmatrix} A_\theta^T & A_\varphi^T & A_\beta^T \end{bmatrix}^T, \quad
\mathbf{b} = \begin{bmatrix} \mathbf{b}_\theta^T & \mathbf{b}_\varphi^T & \mathbf{b}_\beta^T \end{bmatrix}^T, \\
A_\theta, A_\varphi \in \mathbb{R}^{K \times 3}, \quad A_\beta \in \mathbb{R}^{(K-1) \times 3}, \\
A \in \mathbb{R}^{(3K-1) \times 3}, \quad \mathbf{b} \in \mathbb{R}^{(3K-1) \times 1}, \\
A_\theta = \begin{bmatrix} \mathbf{a}_{\theta 1} \\ \vdots \\ \mathbf{a}_{\theta K} \end{bmatrix}, \quad 
A_\varphi = \begin{bmatrix} \mathbf{a}_{\varphi 1} \\ \vdots \\ \mathbf{a}_{\varphi K} \end{bmatrix}, \quad
A_\beta = \begin{bmatrix} \mathbf{a}_{\beta 2} \\ \vdots \\ \mathbf{a}_{\beta K} \end{bmatrix},
\end{gathered}
\label{eq41}
\end{equation}
where:
\begin{equation}
\begin{gathered}
\mathbf{a}_{\theta k} = \begin{bmatrix} \sin{\theta_k^0} & -\cos{\theta_k^0} & 0 \end{bmatrix}, \\
\mathbf{a}_{\varphi k} = \begin{bmatrix} \sin{\varphi_k^0}\cos{\theta_k^0} & \sin{\varphi_k^0}\sin{\theta_k^0} & -\cos{\varphi_k^0} \end{bmatrix}, \\
\mathbf{a}_{\beta k} = 
\begin{bmatrix}
\cos{\varphi_k^0} \cos{\theta_k^0} - \beta_{1k}^0 \cos{\varphi_1^0} \cos{\theta_1^0}, \\
\cos{\varphi_k^0} \sin{\theta_k^0} - \beta_{1k}^0 \cos{\varphi_1^0} \sin{\theta_1^0}, \\
\sin{\varphi_k^0} - \beta_{1k}^0 \sin{\varphi_1^0}
\end{bmatrix}^T,
\end{gathered}
\label{eq42}
\end{equation}
and:
\begin{equation}
\begin{gathered}
b_{\theta k} = x_k \sin{\theta_k^0} - y_k \cos{\theta_k^0}, \\
b_{\varphi k} = x_k \sin{\varphi_k^0} \cos{\theta_k^0} + y_k \sin{\varphi_k^0} \sin{\theta_k^0} - z_k \cos{\varphi_k^0}, \\
b_{\beta k} = x_k \cos{\varphi_k^0} \cos{\theta_k^0} + y_k \cos{\varphi_k^0} \sin{\theta_k^0} + z_k \sin{\varphi_k^0} \\
\quad - \beta_{1k}^0 \left[ x_1 \cos{\varphi_1^0} \cos{\theta_1^0} + y_1 \cos{\varphi_1^0} \sin{\theta_1^0} + z_1 \sin{\varphi_1^0} \right].
\end{gathered}
\label{eq43}
\end{equation}

In practice, measurement errors and noise are inevitable. Using first-order linearization, we model the perturbation  \( \boldsymbol{\varepsilon} \) as
\begin{equation}
A \mathbf{x} - \mathbf{b} \approx \boldsymbol{\varepsilon},
\label{eq44}
\end{equation}
where \(A,\mathbf{b}\) keep the same structure but use measured \(\theta_k,\varphi_k\) and RSS ratios. The error vector is \( \boldsymbol{\varepsilon} = \begin{bmatrix} \boldsymbol{\varepsilon}_\theta^T & \boldsymbol{\varepsilon}_\varphi^T & \boldsymbol{\varepsilon}_\beta^T \end{bmatrix}^T \), 
with components due to azimuth, elevation, and RSS (ratio) errors, respectively. Let \( \mathbf{n} = \begin{bmatrix} (n^\theta)^T & (n^\varphi)^T & (n^s)^T \end{bmatrix}^T \in \mathbb{R}^{(3K - 1) \times 1} \). The RSS errors can be expressed as:
\begin{equation}
\beta_{1k}^0 = \beta_{1k} - \beta_{1k} \, \xi \, (n_k^s - n_1^s),
\quad k = 2, \ldots, K,
\label{eq45}
\end{equation}

By linearizing the measurement models, the perturbation components are expressed as:
\begin{equation}
\begin{gathered}
\boldsymbol{\varepsilon}_\theta = C_\theta n^\theta, \quad
\boldsymbol{\varepsilon}_\varphi = \mathrm{blkdiag}(C_{\varphi \theta}, C_\varphi) 
\begin{bmatrix} n^\theta \\ n^\varphi \end{bmatrix}, \\
\boldsymbol{\varepsilon}_\beta = \begin{bmatrix} \varepsilon_{\beta 2}, \dots, \varepsilon_{\beta K} \end{bmatrix}^T,
\end{gathered}
\label{eq46}
\end{equation}
where \( C_\theta \), \( C_{\varphi \theta} \), and \( C_\varphi \in \mathbb{R}^{K \times K} \) are the Jacobians of the first-order angle perturbations, evaluated at the current estimate and the measured angles:
\begin{equation}
\begin{aligned}
&C_\theta = \text{diag} \{ ..., (x - x_k) \cos{\theta_k} + (y - y_k) \sin{\theta_k} , ...\}, \\
&C_{\varphi \theta} = \text{diag} \{..., -(x - x_k) \sin{\varphi_k} \sin{\theta_k} \\ 
&\qquad \qquad \quad + (y - y_k)\sin{\varphi_k} \cos{\theta_k}, ...\},  \\
&C_\varphi = \text{diag} \{..., (x - x_k) \cos{\varphi_k} \cos{\theta_k} \\
&\qquad + (y - y_k) \cos{\varphi_k} \sin{\theta_k}+ (z - z_k) \sin{\varphi_k}, ...\}.
\end{aligned}
\label{eq47}
\end{equation}

Each \(\varepsilon_{\beta k}\) captures the linearized impact of angular and RSS noise on the distance-ratio componen.
\begin{equation}
\begin{aligned}
\varepsilon_{\beta k} 
&= \big[ (y - y_k)\cos{\varphi_k}\cos{\theta_k} - (x - x_k)\cos{\varphi_k}\sin{\theta_k} \big] n_k^{\theta} \\
&\quad + \big[ (z - z_k)\cos{\varphi_k} - (x - x_k)\sin{\varphi_k}\cos{\theta_k} \\
&\quad - (y - y_k)\sin{\varphi_k}\sin{\theta_k} \big] n_k^{\varphi} + \beta_{1k}\,\xi \big[(z - z_1)\sin{\varphi_1} \\
&\quad + (x - x_1)\cos{\varphi_1}\cos{\theta_1} + (y - y_1)\cos{\varphi_1}\sin{\theta_1} \big] \\
&\quad \times (n_k^s - n_1^s) + \beta_{1k} \big[ (x - x_1)\cos{\varphi_1}\sin{\theta_1} \\
&\quad - (y - y_1)\cos{\varphi_1}\cos{\theta_1} \big] n_1^{\theta} + \beta_{1k} \big[ - (z - z_1)\cos{\varphi_1}  \\
&\quad + (x - x_1)\sin{\varphi_1}\cos{\theta_1} + (y - y_1)\sin{\varphi_1}\sin{\theta_1} \big] n_1^{\varphi}
\end{aligned}
\label{eq48}
\end{equation}

The blocks \(C_{\beta\theta}, C_{\beta\varphi}, C_\beta\) are assembled by collecting the coefficients of \(n^\theta,n^\varphi,n^s\) from \eqref{eq48}, so that:
\begin{equation}
\boldsymbol{\varepsilon} = C \mathbf{n}
= \begin{bmatrix}
C_\theta & \mathbf{0} & \mathbf{0} \\
C_{\varphi \theta} & C_\varphi & \mathbf{0} \\
C_{\beta \theta} & C_{\beta \varphi} & C_\beta
\end{bmatrix}
\begin{bmatrix}
n^\theta \\
n^\varphi \\
n^s
\end{bmatrix}.
\label{eq49}
\end{equation}

The fusion center applies the WLS estimator~\cite{bib32}:
\begin{equation}
\hat{\mathbf{x}}
= \bigl(A^T Q^{-1} A\bigr)^{-1}\,A^T\,Q^{-1}\,\mathbf{b},
\label{eq50}
\end{equation}
where $Q = \mathbb{E}[\boldsymbol{\varepsilon}\,\boldsymbol{\varepsilon}^T]
= C\,\mathbb{E}[\mathbf{n}\,\mathbf{n}^T]\,C^T
= C\,W\,C^T$ is the covariance matrix of the error vector, and $W = \mathbb{E}[\mathbf{n}\,\mathbf{n}^T]
= \mathrm{blkdiag}\bigl(\sigma_\theta^2 I,\;\sigma_\varphi^2 I,\;\sigma_{s}^2 I\bigr)$. 

Hence, its computational complexity scales linearly as $\mathcal{O}(N_{\mathrm{meas}})$, where $N_{\mathrm{meas}}$ denotes the total number of available
RSS and AOA measurements.

When measurement noise is sufficiently small, the WLS performance approaches the Cramér–Rao lower bound (CRLB)~\cite{bib17}:
\begin{equation}
\mathrm{CRLB}_{\mathbf{x}} \approx \left( A^T Q^{-1} A \right)^{-1}.
\label{eq51}
\end{equation}

\subsection{NLOS Environment Localization Algorithm}
While the WLS estimator performs well in LOS conditions, its accuracy deteriorates sharply in NLOS environments, where large bias terms and nonlinear propagation effects are not properly captured. To overcome this, we employ a cooperative framework based on the LSRE criterion, which minimizes the relative mismatch between estimated and measured distances. Within this framework, we investigate two worst-case variants: LSRE--SDP and LSRE--SOCP. First, to improve robustness, we assume the NLOS bias lies within a known interval, $b_k \in [0, b_{\max}]$, and apply a worst-case reformulation per link. From Equation~\eqref{eq8}, rearranging and shifting by \( b_{\max}/2 \) yields:
\begin{equation}
    P_{t|\mathrm{dB}} - P_{r,k|\mathrm{dB}} - \frac{b_{\text{max}}}{2} =  b_k  - \frac{b_{\text{max}}}{2} + 10\beta_k log_{10} (d_k)- n_k^s,
\label{eq52}
\end{equation}

Define $\hat{b}_k = b_k-\frac{b_{\text{max}}}{2}$ and $\hat{P}_{r,k|\mathrm{dB}} = P_{r,k|\mathrm{dB}}+\frac{b_{\text{max}}}{2}$, then:
\begin{equation}
P_{t|\mathrm{dB}} - \hat{P}_{r,k|\mathrm{dB}} = \hat{b}_k + 10 \beta_k \log_{10} (d_k) - n_k^s.
\label{eq53}
\end{equation}
where $0 \leq b_k \leq b_{max}$ and $|\hat{b}_k| \leq \frac{b_{\text{max}}}{2}$.

Exponentiating both sides and grouping terms gives:
\begin{equation}
\frac{m_t}{m_k} = d_k e_k \eta_k , \quad k = 1, \dots, K,
\label{eq54}
\end{equation}
where \( m_t = 10^{\frac{P_{t|\mathrm{dB}}}{10\beta_k}} \), \( m_k = 10^{\frac{\hat{P}_{r,k|\mathrm{dB}}}{10\beta_k}} \), \( d_k = \| \mathbf{a}_k - \mathbf{x} \| \), \( e_k = 10^{\frac{\hat{b}_k}{10\beta_k}} \), and \( \eta_k = 10^{-\frac{n_k^s}{10\beta_k}} \).

A symmetric LSRE loss per link is:
\begin{equation}
\ell_k \;=\; 
\Big(\tfrac{\frac{m_t}{m_k}}{d_k e_k} - 1\Big)^2
+ \Big(\tfrac{d_k e_k}{\frac{m_t}{m_k}} - 1\Big)^2 .
\label{eq55}
\end{equation}

Let $x_k = \tfrac{m_t}{m_k d_k e_k} > 0$. Since 
$(x_k-1)^2 + (x_k^{-1}-1)^2 = x_k^2 + x_k^{-2} - 2$,
minimizing $\sum_k \ell_k$ is equivalent (up to a constant) to minimizing:
\begin{equation}
\hat{\mathbf{x}} = \arg\min_{\mathbf{x}} 
\sum_{k=1}^K \left(
\frac{m_t^2}{m_k^2 d_k^2 e_k^2} + \frac{m_k^2 d_k^2 e_k^2}{m_t^2}
\right),
\label{eq56}
\end{equation}
which is log-convex in the positive variables $(d_k e_k)$ and $m_t$ and can be a convex program using substitutions.

To account for the worst-case NLOS bias, we maximize the impact of $e_k$:
\begin{equation}
    \min_{\mathbf{x}} \max_{e_k} \sum_{k=1}^K \left( \frac{m_t^2}{m_k^2 d_k^2 e_k^2} + \frac{m_k^2 d_k^2 e_k^2}{m_t^2} \right),
    \label{eq57}
\end{equation}

From $\hat{b}_k \in [-\tfrac{b_{\max}}{2},\, \tfrac{b_{\max}}{2}]$, then $e_k \in [\underline e_k,\,\overline e_k], \underline e_k = 10^{-\frac{b_{\max}}{20 \beta_k}}, \overline e_k = \underline e_k^{-1}.$
For each $k$, the function 
$\phi_k(e) \!=\! \frac{m_t^2}{m_k^2 d_k^2 e^2} + \frac{m_k^2 d_k^2 e^2}{m_t^2}$
is convex in $e^2$ and attains its maximum over $[\underline e_k,\,\overline e_k]$ at an endpoint. Hence, the conservative upper bound to the robust objective is:
\begin{equation}
\min_{\mathbf{x}} \;\sum_{k=1}^K \left\{
\frac{m_t^2}{m_k^2 d_k^2 \underline e_k^{2}} + \frac{m_k^2 d_k^2 \underline e_k^{2}}{m_t^2} + 
\frac{m_t^2}{m_k^2 d_k^2 \overline e_k^{2}} + \frac{m_k^2 d_k^2 \overline e_k^{2}}{m_t^2}
\right\} .
\label{eq58}
\end{equation}

Since the objective function in Equation~\eqref{eq58} is non-convex and nonlinear, direct solution is challenging. We apply a variable substitution $\zeta= \frac{1}{m_t}>0$, yielding:
\begin{align}
\min_{\chi,\zeta} \; \sum_{k=1}^K \bigg(
& \frac{1}{\zeta^2 m_k^2 d_k^2 \underline e_k^2} + \zeta^2 m_k^2 d_k^2 \underline e_k^2 \nonumber \quad \quad \\
&  \quad \quad  + \frac{1}{\zeta^2 m_k^2 d_k^2 \overline e_k^2} + \zeta^2 m_k^2 d_k^2 \overline e_k^2 \bigg),
\label{eq59}
\end{align}

Let $\chi = \zeta \mathbf{x}$, so $\zeta d_k = \zeta \left\| \mathbf{a}_k - \mathbf{x} \right\|= \left\| \zeta \mathbf{a}_k - \chi \right\|$. The objective function becomes:
\begin{equation}
\begin{aligned}
    \min_{\chi,\zeta} \sum_{k=1}^K \bigg(
    & \frac{1}{\left\| \zeta \mathbf{a}_k - \chi \right\|^2 m_k^2 \underline e_k^2} + \left\| \zeta \mathbf{a}_k - \chi \right\|^2 m_k^2 \underline e_k^2 \\
    & + \frac{1}{\left\| \zeta \mathbf{a}_k - \chi \right\|^2 m_k^2 \overline e_k^2} + \left\| \zeta \mathbf{a}_k - \chi \right\|^2 m_k^2 \overline e_k^2 \bigg) .
\end{aligned}
    \label{eq60}
\end{equation}

To cast Equation~\eqref{eq60} into a convex form, four auxiliary variables \( t_{1k} \), \( t_{2k} \), \( t_{3k} \), and \( t_{4k} \) are introduced to bound the reciprocal and quadratic terms to form an SDP problem~\cite{bib23}:
\begin{equation}
\begin{aligned}
\min_{\chi,\zeta, \{t_{1k}, t_{2k}, t_{3k}, t_{4k}\}} & \sum_{k=1}^K (t_{1k} + t_{2k} + t_{3k} + t_{4k}) \\
\textrm{s.t.} \quad & \dfrac{1}{\left\| \zeta \mathbf{a}_k - \chi \right\|^2 m_k^2 \underline e_k^2} \leq t_{1k}, \\
        & \left\| \zeta \mathbf{a}_k - \chi \right\|^2 m_k^2 \underline e_k^2 \leq t_{2k}, \\
        & \dfrac{1}{\left\| \zeta \mathbf{a}_k - \chi \right\|^2 m_k^2 \overline e_k^2} \leq t_{3k}, \\
        & \left\| \zeta \mathbf{a}_k - \chi \right\|^2 m_k^2 \overline e_k^2 \leq t_{4k}.
\end{aligned}
\label{eq61}
\end{equation}

To ensure convexity, a matrix lifting is applied by defining \( z = [\chi^T, \zeta]^T \in \mathbb{R}^4 \) and the positive semidefinite matrix \( Z = zz^T\). Then, \( \| \zeta a_k - \chi \|^2 \) = \( \mathrm{tr}(C_k Z) \geq 0 \), where \( C_k = \begin{bmatrix} I_3 & -\mathbf{a}_k \\ -\mathbf{a}_k^T & \| \mathbf{a}_k \|^2 \end{bmatrix} \). The resulting SDP becomes~\cite{bib23}:
\begin{equation}
\begin{aligned}
\min_{Z \succeq 0, \{t_{1k}, t_{2k}, t_{3k}, t_{4k}\}} & \sum_{k=1}^K (t_{1k} + t_{2k} + t_{3k} + t_{4k}) \\
\textrm{s.t.} \quad & 
\begin{bmatrix} 
        \text{tr}(C_k Z) & \frac{1}{m_k \underline e_k} \\
        \frac{1}{m_k \underline e_k} & t_{1k} 
    \end{bmatrix} \succeq 0, \\
        & \text{tr}(C_k Z) m_k^2 \underline e_k^2 \leq t_{2k}, \\
        & 
\begin{bmatrix} 
        \text{tr}(C_k Z) & \frac{1}{m_k \overline e_k} \\
        \frac{1}{m_k \overline e_k} & t_{3k} 
    \end{bmatrix} \succeq 0, \\
        & \text{tr}(C_k Z) m_k^2 \overline e_k^2 \leq t_{4k}. \\
\end{aligned}
\label{eq62}
\end{equation}
where we use the standard rank‑1 relaxation.

As a result, its worst-case single-run computational complexity scales polynomially with the problem size and is commonly characterized as $\mathcal{O}(N_{lift}^{3.5})$, where $N_{lift}$ denotes the dimension of the lifted semidefinite variable.

Define \( Z^* \) as the optimal solution obtained from the relaxed SDP. The target position is:
\begin{equation} 
\hat{\mathbf{x}} = \frac{Z^*(1 \!:\! 3,4)}{Z^*(4,4)}.
\label{eq63}
\end{equation}

\looseness=-1
Although the SDP formulation remains convex and typically offers high accuracy, its reliance on matrix lifting leads to substantial computational overhead. To improve efficiency, we reformulate the problem as a second-order cone program using rotated second-order cones, which maintain convexity while eliminating the need for matrix lifting. The resulting SOCP formulation provides a more scalable alternative with significantly lower computational cost. The SOCP formulation is:
\begin{equation}
\begin{aligned}
\min_{\chi, \zeta, \{ s_{kr},\; \tau^{(1)}_{kr},\; \tau^{(2)}_{kr} \}} \quad
& \sum_{k=1}^{K} \sum_{r \in \{\underline{e}_k,\; \overline{e}_k\}} \left( \tau^{(1)}_{kr} + \tau^{(2)}_{kr} \right) \\
\text{s.t.} \quad
& \left( \tau^{(1)}_{kr},\; y_{kr},\; \sqrt{2} \right) \in \mathcal{Q}^r, \\
& \left( y_{kr},\; \tfrac{1}{2},\; m_k^r s_{kr} \right) \in \mathcal{Q}^r, \\
& \left( \tau^{(2)}_{kr},\; \tfrac{1}{2},\; m_k^r s_{kr} \right) \in \mathcal{Q}^r, \\
& \left\| \zeta \mathbf{a}_k - \chi \right\| \leq s_{kr}, \quad \forall k,\; r \in \{\underline{e}_k,\; \overline{e}_k\}.
\end{aligned}
\label{eq64}
\end{equation}

In this formulation, each link \( k \) and endpoint \( r \in \{ \underline{e}_k, \overline{e}_k \} \) is associated with three auxiliary variables: \( s_{kr} \), \( \tau^{(1)}_{kr} \), and \( \tau^{(2)}_{kr} \), representing the scaled distance, reciprocal term, and quadratic term, respectively. The reciprocal term is modeled using a pair of rotated second-order cones:
\[
\left( \tau^{(1)}_{kr},\; y_{kr},\; \sqrt{2} \right) \in \mathcal{Q}^r,
\quad
\left( y_{kr},\; \tfrac{1}{2},\; m_k^r s_{kr} \right) \in \mathcal{Q}^r,
\]
while the quadratic term is captured by:
\[
\left( \tau^{(2)}_{kr},\; \tfrac{1}{2},\; m_k^r s_{kr} \right) \in \mathcal{Q}^r.
\]

Therefore, its worst-case single-run computational complexity scales polynomially with the problem size and is approximately given by $\mathcal{O}(N_{opt}^{3})$, where $N_{opt}$ denotes the number of optimization variables involved in the SOCP formulation.

From the SOCP solution to Equation~\eqref{eq64}, the target position is recovered as:
\begin{equation}
\hat{\mathbf{x}} = \frac{\hat{\chi}}{\hat{\zeta}},
\label{eq65}
\end{equation}
where \( \hat{\chi} \) and \( \hat{\zeta} \) denote the optimal solution variables.

Both the SDP and SOCP formulations are convex and can be solved efficiently using standard conic solvers such as MOSEK, SeDuMi, or SDPT3. However, these formulation implicitly assumes that the transmit power and PLE are known and fixed, which limits their practicality. In real post-disaster environments, these parameters are unknown and vary across UAV–target links. Thus, we introduce an iterative refinement procedure that jointly estimates the target position, transmit power, and path-loss exponent, extending the LSRE framework to more realistic and dynamic conditions.

Following standard robust localization practice~\cite{bib19}, we initialize the algorithm using empirical estimates of the transmit power \( P_t \) and the path loss exponent parameter \( \beta_k \) for all UAVs, as suggested in~\cite{bib29}. With these initial values, we solve the SOCP problem in Equation~\eqref{eq64} to obtain the optimal variables \( \hat{\chi}^{(j)} \) and \( \hat{\zeta}^{(j)} \). At iteration \( j \), the estimated target position is $\hat{\mathbf{x}}^{(j)} = \frac{\hat{\chi}^{(j)}}{\hat{\zeta}^{(j)}}$.

Recomputing the distance between the target and the \( k \)-th UAV with updated position as:
\begin{equation}
\hat{d}_k^{(j)} = \left\| \hat{\mathbf{x}}^{(j)} - \mathbf{a}_k \right\|.
\label{eq66}
\end{equation}

\looseness=-1
The transmit power is then refined by averaging the reconstructed RSS values from all UAVs based on the updated distances:
\begin{equation}
\hat{P}_{t|\mathrm{dB}}^{(j)} = \frac{1}{K} \sum_{k=1}^{K} \big( P_{r,k|\mathrm{dB}} + 10 \hat{\beta}_k^{(j-1)} \log_{10} ( \hat{d}_k^{(j)}) \big).
\label{eq67}
\end{equation}

The path-loss exponent for each UAV–target link is then updated by inverting the RSS equation:
\begin{equation}
\hat{\beta}_k^{(j)} = \frac{\hat{P}_{t|\mathrm{dB}}^{(j)} - P_{r,k|\mathrm{dB}}}{10 \log_{10} \left( \hat{d}_k^{(j)} \right)}.
\label{eq68}
\end{equation}

This update cycle is repeated iteratively until one of the following stopping criteria is met:
\begin{itemize}
    \item Position Convergence: \( \left\| \hat{\mathbf{x}}^{(j)} - \hat{\mathbf{x}}^{(j-1)} \right\| < \epsilon_{\mathbf{x}} \),
    \item Parameter Convergence: \( \left| \hat{P}_t^{(j)} - \hat{P}_t^{(j-1)} \right| < \epsilon_{P_t} \) and \( \left\| \hat{\beta}_k^{(j)} - \hat{\beta}_k^{(j-1)} \right\| < \epsilon_{PLE} \),
    \item The maximum number of iterations \( J_{\max} \) is reached,
\end{itemize}
where \( \epsilon_{\mathbf{x}} \), \( \epsilon_{P_t} \), and \( \epsilon_{PLE} \) are predefined thresholds.

To ensure stability, we monitor the squared error of the estimated RSS values:
\begin{equation}
J^{(j)} = \sum_{k=1}^K \left( P_{r,k|\mathrm{dB}} - \hat{P}_{t|\mathrm{dB}}^{(j)} + 10 \hat{\beta}_k^{(j)} \log_{10} \left( \hat{d}_k^{(j)} \right) \right)^2.
\label{eq69}
\end{equation}

If \( J^{(j)} \) increases steadily across iterations, the algorithm is considered to diverge and is terminated early. Otherwise, a limited increase over a few iterations is tolerated. To further mitigate divergence, a relaxation mechanism is adopted by introducing a step size \( \nu \in (0,1] \) to update the transmit power:
\begin{equation}
\hat{P}_{t|\mathrm{dB}}^{(j)} \leftarrow \hat{P}_{t|\mathrm{dB}}^{(j-1)} + \nu \left( \hat{P}_{t|\mathrm{dB}}^{(j)} - \hat{P}_{t|\mathrm{dB}}^{(j-1)} \right).
\label{eq70}
\end{equation}

This strategy smooths out fluctuations in the parameter updates, enhancing convergence stability.

\subsection{Mixed Environment Localization Algorithm}

In mixed propagation environments, exactly one UAV maintains a LOS link to the target, while the remaining UAVs experience NLOS propagation. This subsection presents a localization algorithm that exploits the reliable LOS AOA measurement to geometrically constrain the feasible solution space, while simultaneously incorporating both LOS and NLOS RSS observations via a channel‑aware LSRE–SOCP framework consistent with Section~IV‑B.

Let the LOS UAV be located at \( \mathbf{a}_1 = [x_1, y_1, z_1]^T \), which observes the target with measured azimuth and elevation angles \( \theta_1 \) and \( \varphi_1 \). LOS unit direction vector is:
\begin{equation}
\boldsymbol{\delta}_1 =
\begin{bmatrix}
\cos \varphi_1 \cos \theta_1 \\
\cos \varphi_1 \sin \theta_1 \\
\sin \varphi_1
\end{bmatrix}.
\label{eq71}
\end{equation}

All UAVs collect RSS measurements \( \{ P_{r,k|\mathrm{dB}} \} \), and a binary label vector \( \mathbf{y} = [y_1, y_2, \dots, y_K]^T \in \{0,1\}^K \) indicates LOS (\( y_k = 1 \)) or NLOS (\( y_k = 0 \)) links. According to the classifier, we assume that UAV~1 is the unique LOS anchor (\(y_1=1\)) and all other links are NLOS.

We adopt the same robust LSRE–SOCP construction as in \eqref{eq64}, with the following mixed‑environment specialization:
(i) for the LOS link, we set \(b_1=0\) so that \(e_1=1\) and use a fixed path‑loss exponent \(\beta_{\mathrm{LOS}}\);
(ii) for each NLOS link, we keep the per‑link parameters and bias interval \(e_k\in[\underline e_k,\overline e_k]\) as in \eqref{eq58}.
Define endpoint set:
\begin{equation}
\mathcal{R}_k \,=\, 
\begin{cases}
\{1\}, & y_k=1 \text{ (LOS)},\\
\{\underline e_k,\,\overline e_k\}, & y_k=0 \text{ (NLOS)}.
\end{cases}
\label{eq72}
\end{equation}

Then the mixed LSRE objective is:
\begin{equation}
\min_{\chi,\zeta,\,\{s_{kr},y_{kr},\tau^{(1)}_{kr},\tau^{(2)}_{kr}\}}
\ \sum_{k=1}^{K}\ \sum_{r\in\mathcal{R}_k}
\big(\tau^{(1)}_{kr} + \tau^{(2)}_{kr}\big),
\label{eq73}
\end{equation}
subject to the rotated second-order cones constraints reused from \eqref{eq64} for every \(k\) and \(r\in\mathcal{R}_k\). Solving \eqref{eq73} yields an unconstrained target estimate \( \hat{\mathbf{x}} \in \mathbb{R}^3 \).

Because only one high‑confidence LOS direction is available, residual drift along the LOS axis may remain. We enforce geometric consistency by projecting the unconstrained estimate onto the LOS ray from UAV~1:
\begin{equation}
\rho = (\hat{\mathbf{x}} - \mathbf{a}_1)^T \boldsymbol{\delta}_1, 
\qquad 
\tilde{\rho} = \max\{0,\,\rho\},
\label{eq74}
\end{equation}
and the optimized target position is:
\begin{equation}
\hat{\mathbf{x}} = \mathbf{a}_1 + \tilde{\rho}\,\boldsymbol{\delta}_1.
\label{eq75}
\end{equation}

To avoid large errors when the unique LOS link is misclassified, the projection is treated as an optional refinement driven by the RSS fit. We denote by $\hat{\mathbf{x}}_{\mathrm{SOCP}}$ the unconstrained estimate obtained from the mixed LSRE--SOCP problem in \eqref{eq73}, and by $\hat{\mathbf{x}}_{\mathrm{proj}}$ the projected estimate in \eqref{eq75}. Let $J(\cdot)$ be the RSS mismatch cost in \eqref{eq69}, and define
$J_{\mathrm{SOCP}} \triangleq J\big(\hat{\mathbf{x}}_{\mathrm{SOCP}}\big)$ and
$J_{\mathrm{proj}} \triangleq J\big(\hat{\mathbf{x}}_{\mathrm{proj}}\big)$.
For a small tolerance $\epsilon_{\mathrm{proj}}>0$, the target estimate is:
\begin{equation}
\hat{\mathbf{x}} =
\begin{cases}
\hat{\mathbf{x}}_{\mathrm{proj}}, &
    J_{\mathrm{proj}} \le (1+\epsilon_{\mathrm{proj}})\,J_{\mathrm{SOCP}},\\[2pt]
\hat{\mathbf{x}}_{\mathrm{SOCP}}, &
    \text{otherwise.}
\end{cases}
\label{eq76}
\end{equation}
If the link labeled as LOS is in fact NLOS (false positive), the incorrect LOS ray is inconsistent with the RSS geometry, so projecting onto it typically increases $J(\cdot)$; the condition in~\eqref{eq76} then rejects the projected solution and retains $\hat{\mathbf{x}}_{\mathrm{SOCP}}$, so that the performance simply falls back to the underlying LSRE--SOCP baseline, which is designed to be robust in harsh NLOS/mixed environments, even under a mismatched $\beta_{\mathrm{LOS}}$ on the misclassified link.

\begin{algorithm}[h]
\caption{SAR Adaptive Localization Framework}
\begin{algorithmic}[1]
\STATE \textbf{Input:} Averaged RSS and AOA measurements from $K \geq 4$ UAVs.
\STATE \textbf{Step 1: Data Collection} \\
Collect per-link RSS and AOA observations from the UAV swarm.
\STATE \textbf{Step 2: Channel Classification} \\
Apply the CNN–LSTM model to assign binary LOS/NLOS labels to each UAV–target link.
\STATE \textbf{Step 3: Environment Inference}
\IF{Number of LOS links $\ge 2$}
  \STATE Assign \textbf{LOS environment}.
\ELSIF{Number of LOS links $= 1$}
  \STATE Assign \textbf{Mixed environment}.
\ELSE
  \STATE Assign \textbf{NLOS environment}.
\ENDIF
\STATE \textbf{Step 4: Localization Algorithm Selection}
\IF{LOS environment}
    \STATE Apply joint RSS/AOA Taylor-WLS localization.
\ELSIF{NLOS environment}
    \STATE Apply iterative LSRE–SOCP to jointly estimate $\hat{\mathbf{x}}, P_t, \beta_k$.
\ELSIF{Mixed environment}
    \STATE Solve LSRE--SOCP problem to obtain $\hat{\mathbf{x}}_{\mathrm{SOCP}}$.
    \STATE Project $\hat{\mathbf{x}}_{\mathrm{SOCP}}$ onto the LOS ray to obtain $\hat{\mathbf{x}}_{\mathrm{proj}}$.
    \STATE Compute $J_{\mathrm{SOCP}} $ and $J_{\mathrm{proj}}$.
    \IF{$J_{\mathrm{proj}} \le (1+\epsilon_{\mathrm{proj}}) J_{\mathrm{SOCP}}$}
        \STATE Set $\hat{\mathbf{x}} = \hat{\mathbf{x}}_{\mathrm{proj}}$.
    \ELSE
        \STATE Set $\hat{\mathbf{x}} = \hat{\mathbf{x}}_{\mathrm{SOCP}}$.
    \ENDIF
\ENDIF
\STATE \textbf{Output:} Estimated 3D target location $\hat{\mathbf{x}}$.
\end{algorithmic}
\label{Algorithm1}
\end{algorithm}

\section{Discussion}
All runtime and inference evaluations are conducted on a consumer-grade desktop equipped with an Intel\textsuperscript{\textregistered} Core\textsuperscript{TM} i7-11700 CPU (8 cores @ 2.50 GHz) and 16 GB of RAM. Although this hardware is not optimized for high-performance computing, our proposed framework still achieves competitive execution times, underscoring its suitability for field-ready systems and potential deployment on embedded platforms. The computational results reported in Table~\ref{tabIII} and Table~\ref{tabIV}  reflect the actual elapsed time using Python-based implementations (PyTorch and scikit-learn) and MATLAB R2025a. The complete NYUSIM-based dataset used in this study is publicly available on IEEE Dataport~\cite{bib37}.

\subsection{Dataset Generation}
\looseness=-1
To generate realistic LOS/NLOS-labeled training data, we use the NYUSIM v4.0 ~\cite{bib34}~\cite{bib35} channel simulator under UMa (Urban Macrocell) scenarios. The simulation operates at 867.5 MHz with a 3 MHz bandwidth to emulate the NPSPAC band. In total, we generate 41,000 samples, distributed such that 50\% fall within 100\,m, 30\% within 100–150\,m, and the remaining 20\% within 150–200\,m. The LOS/NLOS split is balanced.

We adopt a distance range of 10--500\,m and fix the transmit power uniformly between 13 and 18\,dBm. In NYUSIM, the base station is elevated and the mobile user moves on the ground. To match our UAV-target framework, we invert the XYZ coordinates so the target (base station) is always at \([0, 0, 0]\), and UAV positions are generated dynamically with user heights in the 0--20\,m range. Foliage loss and outdoor-to-indoor penetration loss are sampled according to the LOS/NLOS label.

\looseness=-1
The antenna pattern is configured for a \(360^{\circ}\) azimuth beamwidth and \(45^{\circ}\) elevation beamwidth, supporting omnidirectional reception at the target while enabling directional AOA estimation at UAV-mounted arrays. Both stationary and mobile scenarios are generated by enabling spatial consistency and introducing random human blockage states. This allows the dataset to reflect disaster conditions such as trapped victims (e.g., earthquakes) and fleeing individuals (e.g., wildfires, floods).

Each NYUSIM sample provides 3D receiver positions, multipath parameters (path delays, powers, and phases), and AOA information. To align with realistic sensor limitations, LOS AOA values are recomputed geometrically from UAV-to-target vectors and rounded to the nearest degree. This adjustment is necessary because the AOA values provided by NYUSIM are cluster-level angles, which cannot reliably represent the direct path between transmitter and receiver. NLOS angles are set to zero, as they are unreliable. CIR waveforms are reconstructed using delay and power data, with additive white Gaussian noise at a thermal noise level of \(-174\)\,dBm/Hz applied before power integration. The resulting link budget is combined with transmit power to produce received power values. Finally, each training instance is structured as: [X, Y, Z, Range, LOS, AOA, ZOA, TotalReceivedPower, LinkBudget, CIR0, ..., CIR1015]. This public dataset is intended to facilitate evaluations of future SAR localization systems under realistic wireless propagation.

\subsection{Deep Learning Validation}
To evaluate the classification performance of both traditional ML methods and deep-learning models on the NYUSIM-based dataset, we divide the data using an 80/20 train–test split. The training set is used for model fitting, and the test set is used to compute standard performance metrics including accuracy, precision, recall, F1-score, AUC, and inference time. All models are evaluated under identical conditions for fairness.

Table~\ref{tabIII} summarizes the test performance of three representative machine learning classifiers: RF, GPC, and SVM. These models are trained using handcrafted statistical features extracted from the min–max–normalized CIR magnitude profiles, including the mean, standard deviation, skewness, kurtosis, log-mean, and the Rician $K$-factor. RF achieves the highest test accuracy (0.9519) and precision (0.9448), followed by SVM and GPC. For each ML-trained model, the inference times in Table~\ref{tabIII} account only for the forward-pass evaluation and do not include the offline feature-extraction pipeline used to compute statistical descriptors from the raw CIR data.

In comparison, the proposed CNN–LSTM model is trained directly on raw CIR magnitude sequences, removing the need for manual feature engineering. The convolutional layers extract localized spectral features, while the LSTM layers capture temporal dependencies characteristic of multipath propagation. Despite operating on minimally preprocessed input, the CNN–LSTM achieves a competitive test accuracy of 0.9290, with precision of 0.9108, recall of 0.9502, and F1-score of 0.9301. Moreover, it achieves the highest AUC 0.9696 among all classifiers, demonstrating superior capability in distinguishing between LOS and NLOS channels. The proposed CNN--LSTM architecture contains approximately $7.2 \times 10^{4}$ trainable parameters and requires about $2$--$3$~MFLOPs per inference. 

While RF attains slightly higher raw test accuracy on our current NYUSIM split and enjoys a lower nominal inference time, the CNN--LSTM delivers superior AUC and operates directly on raw CIR data, without the need for manually crafted features. Our goal here is not to propose a novel architecture or to compare CNN and LSTM in depth, as these have already been extensively studied~\cite{bib15,bib16}. Instead, we aim to select a model that is both sufficiently accurate and practical to deploy in realistic SAR settings. The RF is trained using hand-crafted statistical features chosen specifically for the current NYUSIM setup. Porting this design to a new region with different building materials, propagation statistics, or hardware would require additional data collection, nontrivial retraining, and expert intervention, all of which are scarce resources during actual disaster response.

By contrast, the CNN--LSTM is trained offline once on a large and diverse post-disaster dataset and then used as a ready-to-deploy component. During a mission, the UAVs only need to collect standard CIR/RSS measurements and feed them to the pre-trained network; no local ML specialist, feature redesign, or on-site retraining is required. This ``train once, deploy anywhere'' property reduces both the organizational burden on the rescue team and the energy cost of adaptation in the field. In addition, the CNN--LSTM provides the highest AUC (0.9696) among all considered methods, which is particularly important when choosing operating thresholds that heavily penalize dangerous misclassifications (e.g., treating a truly NLOS link as LOS). Taken together, these considerations motivate our use of CNN--LSTM as the default classifier within the proposed framework, while RF remains a strong low-complexity alternative for scenarios with extremely tight on-board resource constraints.

\begin{table}[h]
\caption{Channel Classification Accuracy and Runtime Comparison}
\label{tabIII}
\centering
\begin{tabular}{|l|c|c|c|c|}
\hline
\textbf{Method} & \textbf{Accuracy} & \textbf{Precision} & \textbf{AUC} & \textbf{Time(ms)} \\
\hline
\multicolumn{5}{|c|}{\textit{Machine Learning–Based}} \\
\hline
RF  & 0.9519 & 0.9448 & 0.9499 & 0.0184 \\
GPC & 0.7124 & 0.6920 & 0.7091 & 6.7676 \\
SVM & 0.7001 & 0.6566 & 0.6933 & 0.6435 \\
\hline
\multicolumn{5}{|c|}{\textit{Deep Learning–Based}} \\
\hline
CNN–LSTM & 0.9290 & 0.9108 & \textbf{0.9696} & 0.2544 \\
CNN–LSTM$^{[15]}$ & 0.8156 & -- & -- & -- \\
CNN–LSTM$^{[16]}$ & 0.9653 & -- & 0.9928 & -- \\
\hline
\end{tabular}
\end{table}

\subsection{Analytical Localization Algorithm Validation}
In this subsection, our goal is to isolate and evaluate the localization algorithms themselves. To this end, we consider simplified analytical LOS/NLOS channel models and assume that the LOS/NLOS status of each link is perfectly known from the generative model. The measurements fed to the localization solvers are: (i) AOA, generated from the true geometry and corrupted by Gaussian-distributed angular noise as described in Section II-A, and (ii) RSS, generated from a standard log-normal LOS path-loss model and the empirical log-distance NLOS model in [29] with shadowing and a worst-case bias.

To evaluate the proposed localization algorithm under LOS, Mix, and NLOS environments, we perform extensive Monte Carlo simulations that vary both the number of UAVs and the coverage radius. Localization performance is quantified using the root mean squared error (RMSE):
\begin{equation}
    \mathrm{RMSE} = \sqrt{\mathbb{E}\left[\| \hat{\mathbf{x}} - \mathbf{x} \|^2 \right]},
    \label{eq:rmse}
\end{equation}
where $\hat{\mathbf{x}}$ is the estimated position and $\mathbf{x}$ is the true target location. Each RMSE value is averaged over 5000 independent trials to ensure statistical reliability. 

All simulations are conducted in three dimensions using noisy AOA and RSS measurements. Ground-truth device positions are uniformly sampled within a circular region of radius~$R_0$ and assigned heights between 1~m and 20~m to emulate realistic post-disaster scenarios in which victims may be buried or trapped within multi-level rubble. UAVs operate at a fixed altitude of $H_{\mathrm{UAV}} = 50$~m. When $K = 4$, all UAVs are placed uniformly along the circumference of the outer circle. For $K > 4$, the additional UAVs are placed evenly along an inner ring with reduced radius, maintaining uniform angular spacing while avoiding geometric redundancy. This configuration reflects practical layered UAV layouts commonly used in constrained urban environments to achieve efficient aerial coverage during SAR operations.

\begin{table}[t]
\caption{Methods Comparison of runtime and accuracy}
\centering
\begin{tabular}{|l|c|c|}
\hline
\textbf{Method} & \textbf{Time(s)} & \textbf{RMSE(m)} \\
\hline
\multicolumn{3}{|c|}{\textit{LOS Environment}} \\
\hline
WLS-Ideal & 0.008 & 2.3303 \\
\textbf{WLS-Taylor (Suggested)} & \textbf{0.004 }& \textbf{2.3312} \\
SDP/SOCP~\cite{bib18} & 3.958 & 2.8047 \\
SR-WLS~\cite{bib33} & 0.140 & 3.8335 \\
CRLB & -- & 2.2189 \\
\hline
\multicolumn{3}{|c|}{\textit{NLOS Environment}} \\
\hline
Iterative RWLS-SDP~\cite{bib19} & 3.0590 & 58.7901 \\
Iterative LSRE-SDP & 2.4446 & 61.0825 \\
\textbf{Iterative LSRE-SOCP (Proposed)} & \textbf{2.1455} & \textbf{57.5073} \\
LSRE-SDP~\cite{bib23} & 1.4786 & 79.5862 \\
\hline
\multicolumn{3}{|c|}{\textit{Mixed Environment}} \\
\hline
Iterative RWLS-SDP~\cite{bib19} & 5.0913 & 59.9702 \\
Iterative LSRE-SDP & 1.4955 & 63.8325 \\
Iterative LSRE-SOCP & 2.8710 & 56.7213 \\
1AOA-nRSS~\cite{bib20} & 0.003 & 134.9371 \\
Proj-on-LOS with 3 NLOS RSS & 2.8691 & 36.4277 \\
Angle Reconstruction with WLS-Tay & 2.8646 & 42.6488 \\
\textbf{Proposed Label-aware Proj-on-LOS} & \textbf{1.5092} & \textbf{28.6015} \\
\hline
\multicolumn{3}{|c|}{\textit{Proposed Algorithm Validation Using NYUSIM Data}} \\
\hline
Iterative RWLS-SDP~\cite{bib19} & 4.6570 & 108.1215 \\
Iterative LSRE-SDP & 2.9228 & 69.2648 \\
Iterative LSRE-SOCP & 5.2775 & 66.2640 \\
\textbf{Proposed DL-Based Algorithm} & \textbf{3.0824} & \textbf{46.8314} \\
\hline
\end{tabular}
\label{tabIV}
\end{table}

Two simulation scenarios are designed to evaluate the robustness of the proposed algorithm. In the first scenario, the coverage radius is fixed at $R_0 = 100$~m, and the number of UAVs $K$ is varied from 4 to 7. A minimum of four UAVs is assumed to ensure localization feasibility in NLOS and mixed environments, while additional UAVs are evaluated to assess the benefits of increased geometric diversity. In the second scenario, the number of UAVs is fixed at $K = 4$, and the coverage radius $R_0$ is varied from 100~m to 400~m to examine the impact of increasing distance on localization accuracy.

Each UAV–target link is assigned either a LOS or NLOS channel model depending on its propagation condition. For LOS links, the RSS measurements are modeled using the standard log-distance path loss model with fixed path-loss exponent $\beta = 2.5$ and additive zero-mean Gaussian shadowing noise with a standard deviation of $5$\,dB. For NLOS links, we adopt the empirical channel model from~\cite{bib29} at 900\,MHz using co-polarized antenna configurations. This frequency is based on the tradeoff between penetration capability and distance-dependent attenuation. Specifically, path loss parameters are sampled based on measurements from Scenarios 2, 3, and 4 in Table~II of~\cite{bib29}. In each Monte Carlo trial, one set of path loss exponent, shadow fading, and excess loss is randomly drawn from these empirical ranges. The remaining portion of the UAV-to-target path is modeled using the Friis free-space transmission equation. To emulate variability in beacon strength from battery-constrained mobile devices, the transmit power $P_t$ is randomized uniformly in each trial over the range [13, 18]\,dBm. Furthermore, to reflect unmodeled attenuation due to burial depth, debris density, and unknown scattering, we apply a worst-case NLOS bias bound $b_{\max} \in [6, 8]$\,dB in robust LSRE-based solvers.

\begin{figure}[t]
\centering
\includegraphics[width=1\linewidth]{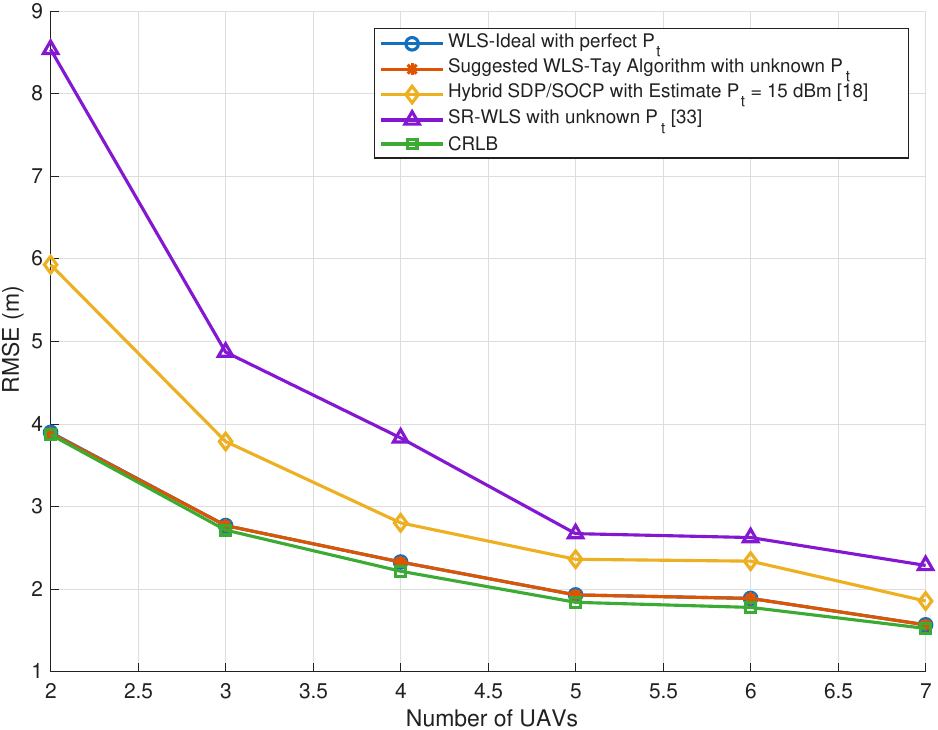}
\caption{LOS $\sigma_\theta=\sigma_\varphi=1^\circ$: RMSE versus the number of UAVs with $R_0=100\,m$.}  
\label{Fig.3}			
\end{figure}

\begin{figure}[t]
\centering
\includegraphics[width=1\linewidth]{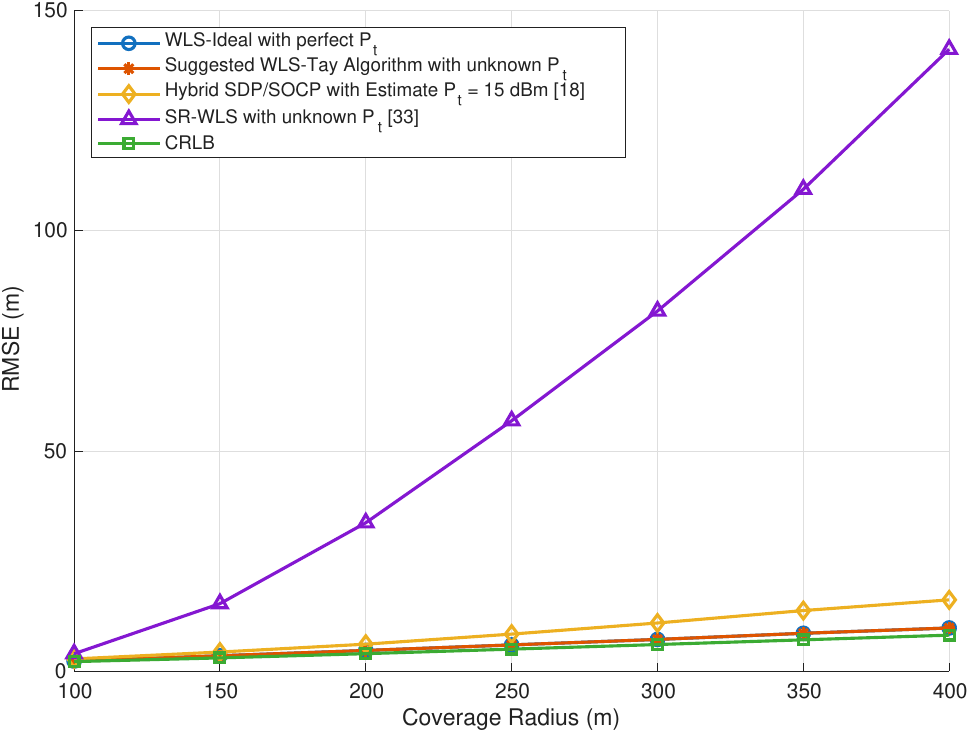}
\caption{LOS $\sigma_\theta=\sigma_\varphi=1^\circ$: RMSE versus coverage radius with $K = 4$ UAVs.}  
\label{Fig.4}			
\end{figure}

\textbf{LOS Environment} is dominated by LOS propagation, where at least two UAV–target links maintain clear line-of-sight conditions while the remaining links may experience mild obstruction. Both RSS and AOA measurements are used for localization, with AOA noise modeled with a standard deviation of $1^\circ$. To evaluate localization performance, we compare our suggested WLS-Taylor algorithm against several representative baselines: the WLS-Ideal baseline, the SDP/SOCP hybrid estimator~\cite{bib18}, the significance regression WLS (SR-WLS) method~\cite{bib33}, and derived CRLB. The resulting RMSE trends are shown in Fig.~\ref{Fig.3} and Fig.~\ref{Fig.4}. Table~\ref{tabIV} summarizes the mean runtime and RMSE for the LOS case with four UAVs and a coverage radius of $R_0 = 100$\,m.

WLS-Ideal assumes perfect knowledge of both AOA and RSS measurements, as well as the true transmit power $P_t$. We also compare against the SDP/SOCP hybrid estimator. While accurate, this method requires the true $P_t$ to be known, which is impractical in post-disaster settings. In our implementation, we assign a fixed $\hat{P}_t = 15$\,dBm to approximate mobile transmit power. This increases localization error and incurs a higher computational costs. In contrast, the suggested WLS-Taylor algorithm does not require $P_t$ and achieves localization accuracy nearly WLS-Ideal, approaching the theoretical performance predicted by CRLB. Lastly, although SR-WLS method also avoids the need for $P_t$, it is highly sensitive to measurement noise and delivers the poorest accuracy.

\begin{figure}[t]
\centering
\includegraphics[width=1\linewidth]{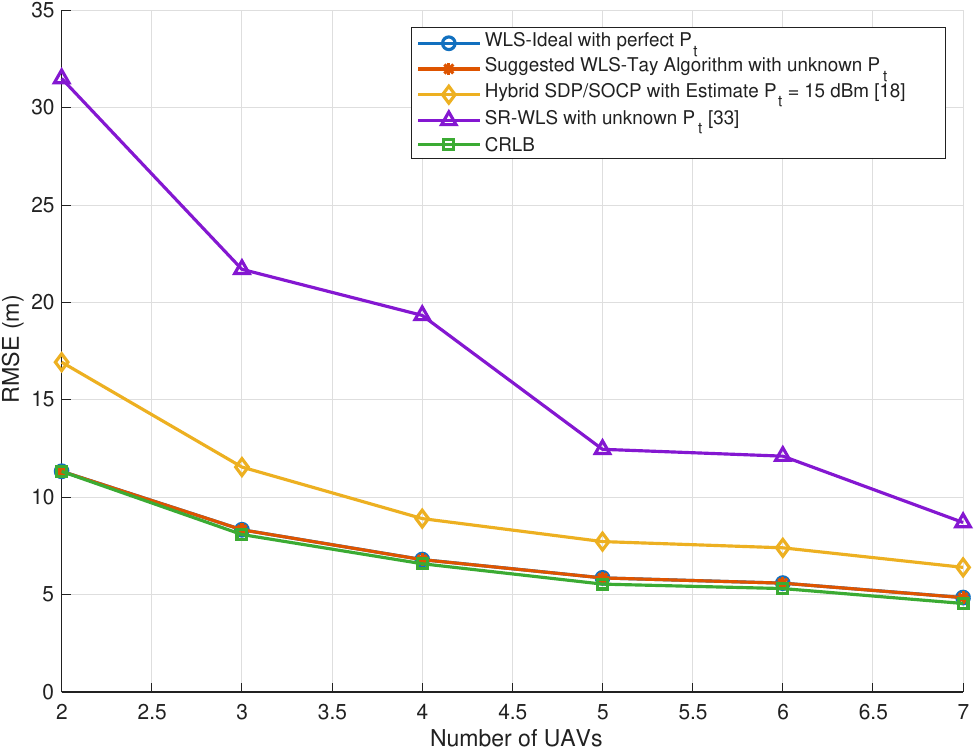}
\caption{LOS $\sigma_\theta=\sigma_\varphi=3^\circ$: RMSE versus the number of UAVs with $R_0=100\,m$.}  
\label{Fig.5}			
\end{figure}

\begin{figure}[t]
\centering
\includegraphics[width=1\linewidth]{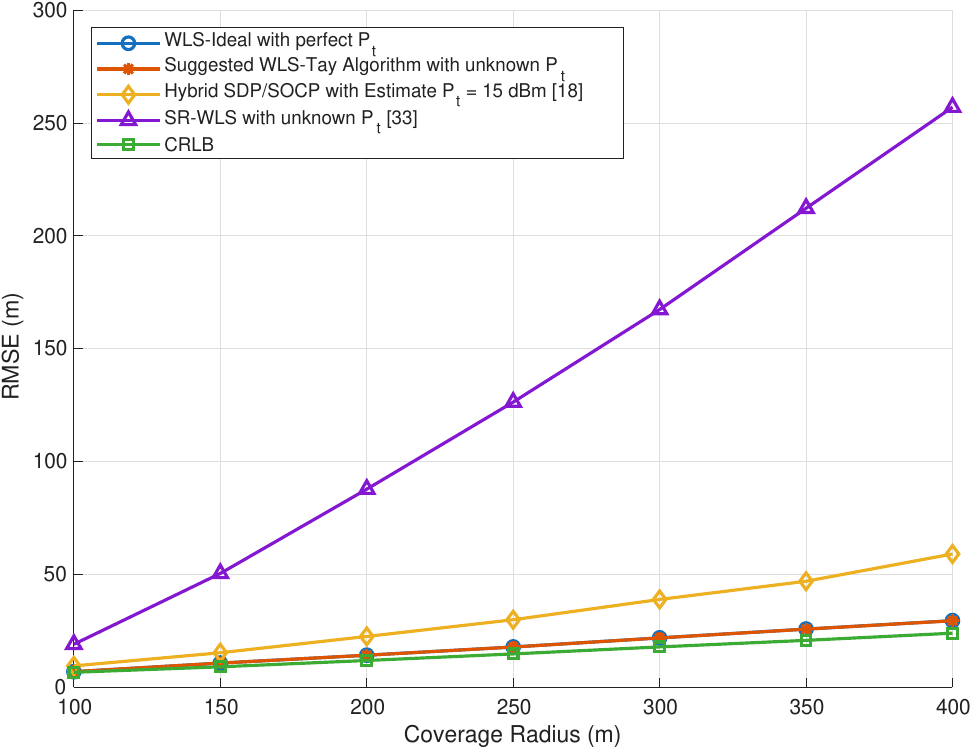}
\caption{LOS $\sigma_\theta=\sigma_\varphi=3^\circ$: RMSE versus coverage radius with $K = 4$ UAVs.}  
\label{Fig.6}			
\end{figure}

To quantify AOA sensitivity, Figs.~\ref{Fig.5} and~\ref{Fig.6} report LOS localization with AOA noise standard deviation increased from $1^\circ$ to $3^\circ$. The RMSE rises as expected, particularly for large $R_0$, but the WLS--Taylor algorithm still stays close to the CRLB and continues to outperform SR--WLS and the SDP/SOCP baseline across all radii. Even under $3^\circ$ AOA error, its LOS error remains significantly smaller than the NLOS and mixed-environment, showing that exploiting available LOS bearings is an effective way to boost localization accuracy in SAR scenarios.

\textbf{NLOS Environment} only has NLOS propagation, where all UAV–target links are obstructed, resulting in severe attenuation, scattering, and unpredictable multipath effects. Only RSS measurements are available for localization, and no angular information is assumed. We evaluate four representative algorithms: Iterative RWLS-SDP~\cite{bib19}, Iterative LSRE-SDP, Proposed iterative LSRE-SOCP, and LSRE-SOCP with estimated $P_t$ and PLE~\cite{bib23}. For LSRE-SDP, we adopt fixed parameter estimates of $\hat{P}_t = 15$\,dBm and $\hat{\beta} = 2.7$.

\looseness=-1
LSRE-SDP achieves the lowest accuracy, and even with iterative refinement it still lags behind the RWLS-SDP benchmark. Here, “iterative” refers to schemes that update both transmit power and PLE across iterations to mitigate model mismatch and improve localization accuracy. The proposed SOCP solver significantly lowers complexity and runtime. This efficiency is important for time-critical SAR operations, where faster inference and improved accuracy directly translate into shorter response times and reduced onboard energy consumption.

\begin{figure}[t]
\centering
\includegraphics[width=1\linewidth]{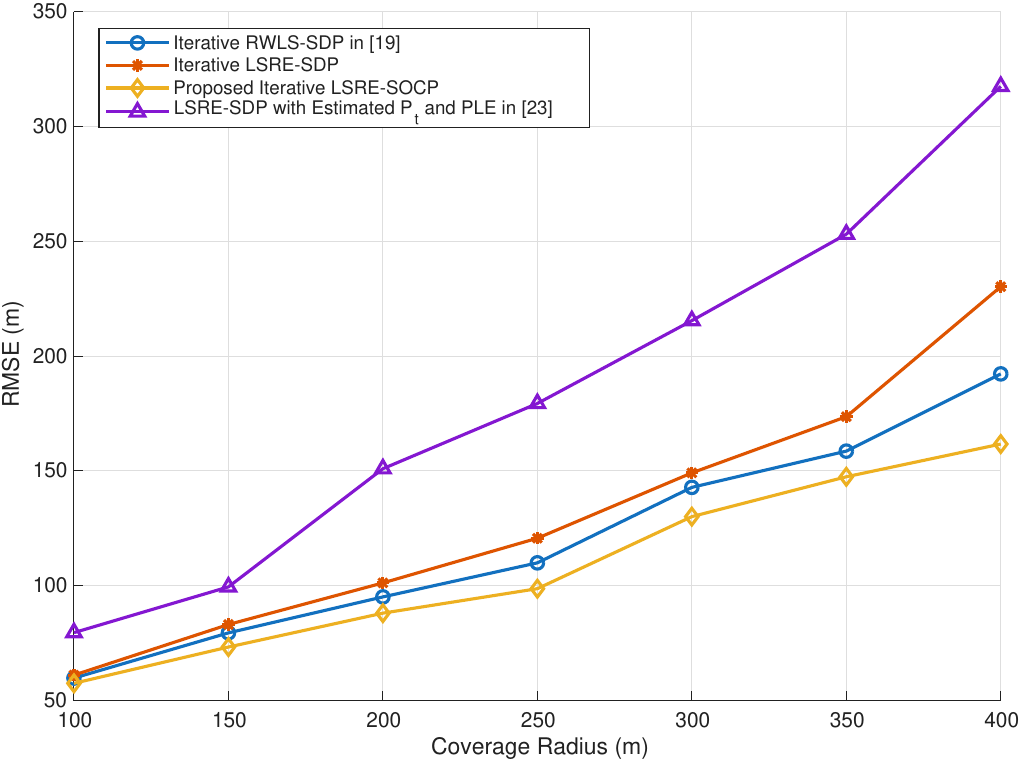}
\caption{NLOS: RMSE versus number of UAVs with $R_0=100\,m$.}  
\label{Fig.7}			
\end{figure}

\begin{figure}[t]
\centering
\includegraphics[width=1\linewidth]{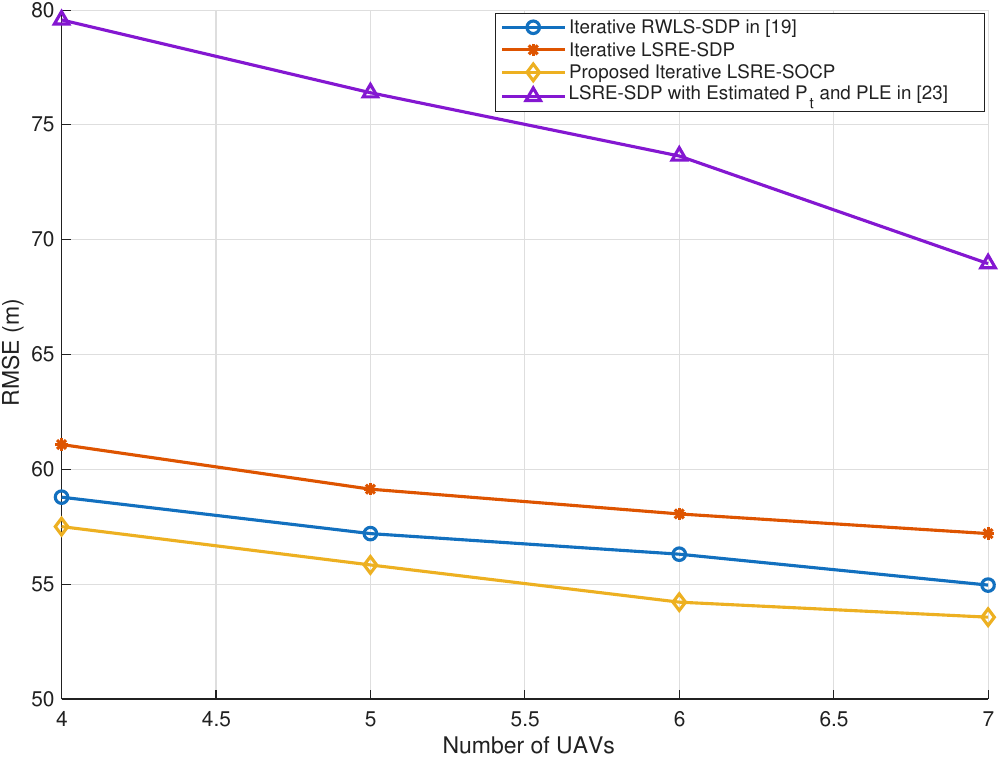}
\caption{NLOS: RMSE versus coverage radius with $K = 4$ UAVs.}  
\label{Fig.8}			
\end{figure}

For $K = 4$ UAVs, the average localization error is approximately 57\,m. Although this RMSE is coarse by GPS standards, it remains operationally valuable in SAR missions. Reducing the search area to a 50–60\,m region within a $200$\,m~$\times$~$200$\,m post-disaster zone ($R_0 = 100$\,m) can effectively accelerate victim discovery. Once responders are guided to the correct region, finer detection tools such as thermal imaging or through-the-wall radar can be used for precise identification. To further constrain the search in 3D, we impose a 20\,m upper bound on the estimated target height, representing typical building collapse conditions and limiting altitude uncertainty. Fig.~\ref{Fig.7} and Fig.~\ref{Fig.8} show the RMSE trends under NLOS conditions as the number of UAVs and the coverage radius vary. Table~\ref{tabIV} listed a comparison of runtime and accuracy.

 \begin{figure}[t]
\centering
\includegraphics[width=1\linewidth]{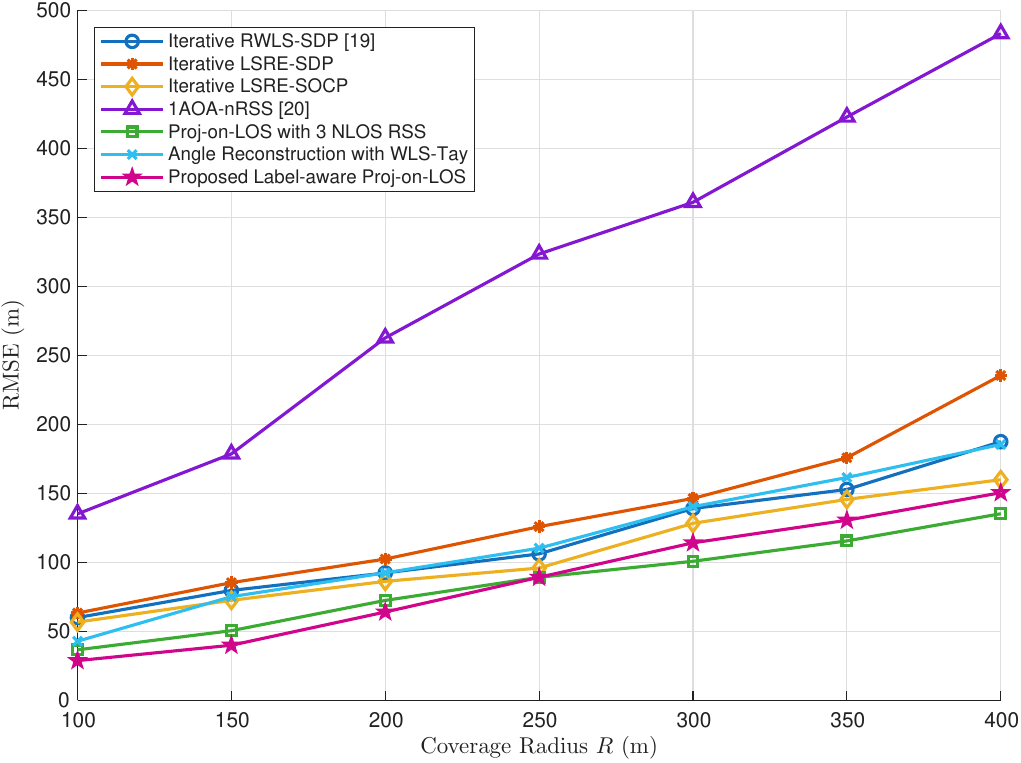}
\caption{Mixed: RMSE versus coverage radius with $K = 4$ UAVs.}  
\label{Fig.9}			
\end{figure}

\looseness=-1
\textbf{Mixed Environment} captures a realistic operating condition in post-disaster SAR missions, where only one UAV maintains a LOS link to the target while the remaining three operate under NLOS conditions. We evaluate seven algorithms, as shown in Fig.~\ref{Fig.9}, with Table~\ref{tabIV} reporting their runtime and RMSE for the mixed environment case with four UAVs. The first three (Iterative RWLS-SDP, Iterative LSRE-SDP, and Iterative LSRE-SOCP) are direct extensions of the NLOS solvers and serve as baseline reference. The fourth method, 1AOA–nRSS~\cite{bib20}, estimates an average bearing vector from the three NLOS anchors and projects a LOS ray from the LOS anchor to estimate targets. However, due to the poor directional reliability of the NLOS links, this strategy exhibits the worst performance.

\looseness=-1
The fifth method introduces angle reconstruction: we first solve the LSRE-SOCP using three NLOS RSS measurements to jointly estimate $P_t$, $\beta$, and distances. We then use this estimated position to geometrically reconstruct the missing NLOS AOA angles following the decomposition in~\cite{bib36}. These reconstructed angles, together with the estimated distances and known RSS noise variances, are fed into the LOS WLS-Taylor solver. While this improves angular coverage, the final accuracy is limited by the quality of the initial LSRE–SOCP estimate.

To improve accuracy, we introduce a projection-based strategy that leverages the reliable LOS AOA measurement to anchor the final estimate. Unlike angle-reconstruction methods that derive NLOS angles from noisy RSS-based distances, our label-aware projection approach sidesteps angular drift by grounding the solution along a trusted LOS direction. This leads to improved stability, faster convergence, and minimal parameter tuning. The first variant, \textit{Proj-on-LOS}, applies LSRE–SOCP to the three NLOS RSS measurements and then projects the resulting estimate onto the LOS AOA ray. This method benefits from the high directional reliability of the LOS angle (with only $1^\circ$ noise) and achieves strong performance across a wide range of distances.

The second variant (\textit{Label-aware Proj-on-LOS}) further incorporates the LOS UAV's RSS measurements, assuming correct channel identification from the classifier. For the LOS link, we fix its PLE to $\beta = 2.5$. All four RSS measurements are used within a label-aware LSRE–SOCP solver, after which the estimate is projected along the LOS AOA direction. Under $K=4$ and $R_0 = 100\,\mathrm{m}$, this reduces the average RMSE from $56.7\,\mathrm{m}$ (Iterative LSRE--SOCP) to $28.6\,\mathrm{m}$ (a nearly $50\%$ improvement), and cuts runtime from $2.87\,\mathrm{s}$ to $1.51\,\mathrm{s}$ (about $47\%$ reduction). Compared with Iterative RWLS--SDP, the proposed approach further decreases RMSE by roughly $52\%$ and runtime by about $70\%$. At larger coverage radius, performance degrades slightly due to increased projection error and reduced reliability of AOA measurements. Nonetheless, fixing the LOS path-loss exponent effectively regularizes the estimation problem and accelerates convergence.

\looseness=-1
In the mixed-environment experiments, we focus on an idealized setting with high-quality LOS bearings to isolate the algorithmic gain of the label-aware projection strategy. The impact of AOA accuracy has already been quantified in the LOS case (Figs.~\ref{Fig.5}–\ref{Fig.6}); when the LOS bearing becomes unreliable, the consistency test in \eqref{eq76} automatically rejects harmful projections and the estimator falls back to the robust LSRE–SOCP baseline. In realistic deployments, the dominant factor in mixed environments is thus the correctness of LOS/NLOS channel classification rather than small variations in AOA quality, and additional mixed-case angle-sensitivity sweeps would not change the qualitative trends observed in Fig.~\ref{Fig.9}.
 
\subsection{NYUSIM-Based End-to-End Environment-Aware Localization Validation}
In this subsection, we use NYUSIM-generated data to evaluate the complete environment-aware end-to-end localization framework. The CNN–LSTM operates directly on CIR sequences to infer per-link LOS/NLOS labels, and the resulting labels together with RSS measurements are then passed to the environment-adaptive localization algorithms. For each Monte Carlo trial, four links are drawn from the NYUSIM database according to a realistic distribution of $15\%$ LOS, $25\%$ NLOS, and $60\%$ mixed environments, and end-to-end localization is performed. In the NYUSIM-based experiments, LOS AoA values are computed geometrically from the UAV–target positions and quantized to the nearest degree, corresponding to the optimistic $\sigma_\theta = \sigma_\varphi \approx 1^\circ$ benchmark analyzed in Section~V-C. As discussed there, this setting should be interpreted as a best-case upper bound on AoA quality; larger angular errors increase the absolute RMSE but do not change the relative ranking or qualitative behavior of the compared methods, especially in NLOS and mixed environments where RSS plays the dominant role.

\begin{figure}[t]
\centering
\includegraphics[width=1\linewidth]{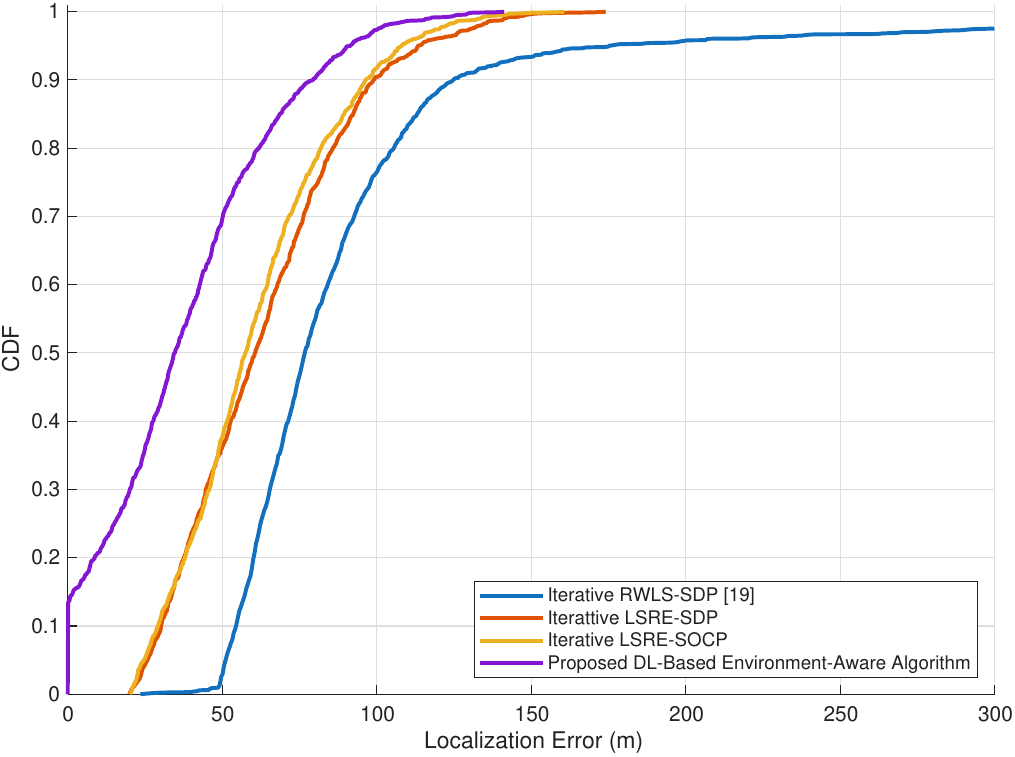}
\caption{CDF of localization error for algorithm validation.}
\label{Fig.10}			
\end{figure}

Fig.~\ref{Fig.10} shows the cumulative distribution function (CDF) of localization error across 5000 trials. Compared to three strong baselines, the proposed framework consistently achieves better localization accuracy. Its CDF curve rises more sharply and saturates earlier, indicating tighter error bounds and greater reliability across heterogeneous conditions. As summarized in Table~\ref{tabIV}, the proposed framework reduces the average RMSE from $66.3\,\mathrm{m}$ (Iterative LSRE--SOCP) to $47.7\,\mathrm{m}$, while also lowering the runtime from $5.28\,\mathrm{s}$ to $3.03\,\mathrm{s}$.

In contrast, RWLS-SDP and LSRE-SDP occasionally produce unreasonable estimates in LOS environments because they lack channel awareness. Without proper initialization of $P_t$ and $\beta$, these methods may diverge and generate outliers exceeding $500$\,m. To maintain fairness, such extreme cases were excluded from evaluation. The SOCP-based baseline is more stable but still incurs higher computational cost and longer runtime.

Although localization accuracy improves substantially, runtime reduction under NYUSIM conditions is smaller than in the analytical channel-model experiments. This is primarily because the fixed LOS path-loss exponent $\beta = 2.5$ contributes only modest acceleration in mixed environments when empirical NYUSIM parameters already dominate the optimization. Even so, the proposed framework remains both efficient and scalable. In real deployments, SAR teams may further refine the LOS $\beta$ through simple field calibration, which can further enhance accuracy and runtime, but such hardware-specific tuning is beyond the scope of this work.

\looseness=-1
The classification labels used in these simulations are provided by the CNN–LSTM model. Although not perfectly accurate, they reflect the expected performance of real-time inference in practical settings. Occasional misclassifications may arise from signal ambiguity or noise; however, the low computational cost of our framework enables simple verification strategies, such as confidence-based filtering or re-execution when abnormal results are detected. Even if the algorithm is rerun for validation, this overhead remains negligible relative to the time, effort, and risk involved in manual SAR missions. Overall, the proposed method significantly narrows the effective search space while maintaining low runtime, making it well suited for practical UAV-assisted SAR deployment.

\section{Conclusion}
The proposed UAV-assisted localization framework achieves robust and accurate 3D positioning across LOS, NLOS, and mixed propagation environments by combining deep-learning-based channel classification with environment-adaptive optimization. Extensive NYUSIM-based simulations validate its efficiency and reliability under realistic post-disaster conditions, demonstrating strong potential for deployment in time-critical SAR missions. Our current prototype assumes a desktop-class fusion node, but the same CNN–LSTM and LSRE–SOCP / Taylor–WLS pipeline can be ported to compact embedded platforms (e.g., NVIDIA Jetson or Raspberry Pi) to build portable UAV-assisted SAR equipment, at the cost of proportionally increased runtime. As future work, we plan to develop and test such embedded implementations in the field and to exploit the resulting accurate localization for higher-level SAR tasks, including motion-aware UAV trajectory planning that uses UAV movement to further refine target positions.

\IEEEtriggeratref{32}

\begin{IEEEbiography}[{\includegraphics[width=1in,height=1.25in,clip,keepaspectratio]{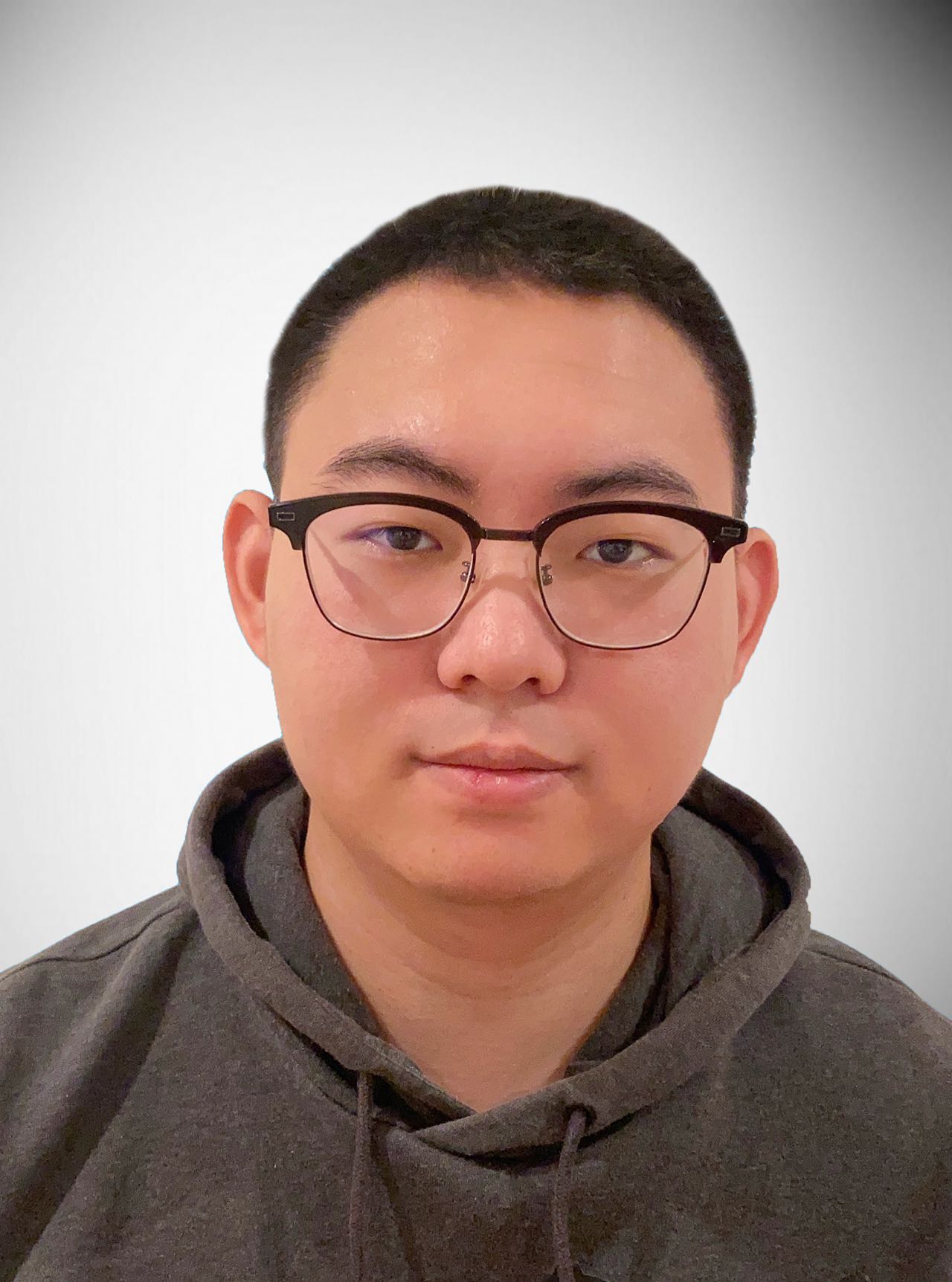}}]{XIANGJIAN GAO} (MEMBER, IEEE) received the B.Sc. degree in electrical engineering from the University of Illinois Urbana-Champaign, Champaign, IL USA in 2016, and the M.Sc. degree in electrical engineering from The George Washington University, Washington, D.C. USA in 2018, and he is currently pursuing the Ph.D. degree in electrical and computer engineering at University of California at Santa Cruz. He received the Best Presenter award at the 11th Electrical Power, Communications, Control, and Informatics Seminar.
\end{IEEEbiography}

\begin{IEEEbiography}[{\includegraphics[width=1in,height=1.25in,clip,keepaspectratio]{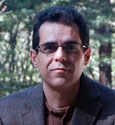}}]{HAMID R. SADJADPOUR} (SENIOR MEMBER, IEEE) received the B.S. and M.S. degrees from the Sharif University of Technology and the Ph.D. degree from the University of Southern California (USC). In December 1995, he joined the ATT Shannon Research Laboratory, as a Technical Staff Member and later as a Principal Member of Technical Staff. In 2001, he joined the University of California at Santa Cruz, where he is currently a Professor. He has authored over 200 publications and holds 25 patents. His research interests are in the general areas of wireless communications, security, and networks. He was a co-recipient of the Best Paper Awards at the 2007 International Symposium on Performance Evaluation of Computer and Telecommunication Systems, the 2008 IEEE Fred W. Ellersick Award in Military Communications conference, the 2010 European Wireless Conference, and the 2017 Conference on Cloud and Big Data Computing. He has served as a technical program committee member and the chair for numerous conferences.
\end{IEEEbiography}

\end{document}